\documentclass[11pt,a4paper]{article}
\usepackage[hyperref]{acl2020}
\usepackage{times}
\usepackage{latexsym}

\usepackage{latexsym}
\usepackage{booktabs}
\usepackage{multirow}
\usepackage{algorithm}
\usepackage{algorithmic}
\usepackage{amsmath}
\usepackage{graphicx}
\usepackage{enumitem}
\usepackage{amssymb}
\usepackage{pifont}
\newcommand\bigforall{\mbox{\Large $\mathsurround0pt\forall$}} 
\usepackage{microtype}

\aclfinalcopy 

\DeclareMathOperator*{\minimize}{min}
\DeclareMathAlphabet\mathbfcal{OMS}{cmsy}{b}{n}

\usepackage{array}
\newcolumntype{H}{>{\setbox0=\hbox\bgroup}c<{\egroup}@{}}

\definecolor{forestgreen}{rgb}{0.0, 0.27, 0.13}
\definecolor{indigo}{rgb}{0.0, 0.25, 0.42}
\definecolor{purple}{rgb}{153, 0, 153}

\usepackage[skip=5pt]{caption}

\newcommand{\mymethod}{\textsf{DARCY}}

\title{A Sweet Rabbit Hole by {\mymethod}:\\Using Honeypots to Detect Universal Trigger's Adversarial Attacks}

\author{Thai Le \\
  Penn State University \\
  \texttt{thaile@psu.edu} \\\And
  Noseong Park \\
  Yonsei University \\
  \texttt{noseong@yonsei.ac.kr}\\\And
  Dongwon Lee \\
  Penn State University \\
  \texttt{dongwon@psu.edu} }
  
\date{}

\begin{document}
\maketitle
\begin{abstract}
The Universal Trigger (\emph{UniTrigger}) is a recently-proposed powerful adversarial textual attack method. Utilizing a learning-based mechanism, UniTrigger generates a fixed phrase that, when added to \textit{any} benign inputs, can drop the prediction accuracy of a textual neural network (NN)  model to near zero on a target class. To defend against this attack that can cause significant harm, in this paper, we borrow the ``honeypot" concept from the cybersecurity community and propose {\mymethod}, a honeypot-based defense framework against UniTrigger. {\mymethod} greedily searches and injects multiple {\em trapdoors} into an NN model to ``bait and catch" potential attacks. Through comprehensive experiments across four public datasets, we show that {\mymethod} detects UniTrigger's adversarial attacks with up to 99\% TPR and less than 2\% FPR in most cases, while maintaining the prediction accuracy (in F1) for clean inputs within a 1\% margin. We also demonstrate that {\mymethod} with multiple trapdoors is also robust to a diverse set of attack scenarios with attackers' varying levels of knowledge and skills. Source code will be released upon the acceptance of this paper.
\end{abstract}

\section{Introduction}
Adversarial examples in NLP refer to carefully crafted texts that can fool predictive machine learning (ML) models. Thus, malicious actors, i.e., attackers, can exploit such adversarial examples to force ML models to output desired predictions. There are several adversarial example generation algorithms, most of which perturb an original text at either character (e.g., \cite{li2018textbugger,gao2018black}), word (e.g., \cite{ebrahimi2017hotflip,jin2019bert,wallace2019universal,gao2018black,garg2020bae}, or sentence level (e.g., \cite{malcom,gan2019improving,cheng2020seq2sick}). Because most of the existing attack methods are instanced-based search methods, i.e., searching an adversarial example for each specific input, and usually do not involve any learning mechanisms. 
A few \textit{learning-based} algorithms, such as the Universal Trigger (\textit{UniTrigger}) \cite{wallace2019universal}, MALCOM \cite{malcom}, Seq2Sick \cite{cheng2020seq2sick} and Paraphrase Network \cite{gan2019improving}, ``learn" to generate adversarial examples that can be effectively generalized to \textit{not a specific} but a wide range of \textit{unseen} inputs.

\begin{table}[t!b]
\centering
\footnotesize
\vspace{-10pt}
\begin{tabular}{ll}
\toprule
\textbf{Original:} & \textit{this movie is awesome}\\
\textbf{Attack:} & \textbf{zoning zoombie} \textit{this movie is awesome}\\
\textbf{Prediction:} & \textcolor{red}{\textbf{Positive}} $\longrightarrow$ \textcolor{blue}{\textbf{Negative}} \\
\midrule
\textbf{Original:} & \textit{this movie is such a waste!} \\
\textbf{Attack:} & \textbf{charming} \textit{this movie is such a waste!} \\
\textbf{Prediction:} & \textcolor{blue}{\textbf{Negative}} $\longrightarrow$ \textcolor{red}{\textbf{Positive}}\\
\bottomrule
\end{tabular}
\caption{Examples of the UniTrigger Attack}
\label{tab:example_UniTrigger}
\vspace{-10pt}
\end{table}

In general, learning-based attacks are more attractive to attackers for several reasons. First, they achieve high attack success rates. For example, UniTrigger can drop the prediction accuracy of an NN model to near zero just by appending a learned adversarial phrase of only two tokens to any inputs (Tables \ref{tab:example_UniTrigger} and \ref{tab:performance_attacks}). This is achieved through an optimization process over an entire dataset, exploiting potential weak points of a model as a whole, not aiming at any specific inputs. Second, their attack mechanism is highly transferable among similar models. To illustrate, both adversarial examples generated by UniTrigger and MALCOM to attack a white-box NN model are also effective in fooling unseen black-box models of different architectures \cite{wallace2019universal,malcom}. Third, thanks to their generalization to unseen inputs, learning-based adversarial generation algorithms can facilitate mass attacks with significantly reduced computational cost compared to instance-based methods. 

Therefore, the task of defending \textit{learning-based} attacks in NLP is critical. Thus, in this paper, we propose a novel approach, named as {\mymethod}, to defend adversarial examples created by UniTrigger, a strong representative learning-based attack (see Sec. \ref{sec:uni}). To do this, we exploit UniTrigger's own advantage, which is the ability to generate a \textit{single} universal adversarial phrase that successfully attacks over several examples. Specifically, we borrow the ``honeypot" concept from \textit{the cybersecurity domain} to bait multiple ``trapdoors" on a textual NN classifier to catch and filter out malicious examples generated by UniTrigger. In other words, we train a target NN model such that it offers great a incentive for its attackers to generate adversarial texts whose behaviors are pre-defined and intended by defenders. Our contributions are as follows:
\begin{itemize}[leftmargin=\dimexpr\parindent-0.2\labelwidth\relax,noitemsep]
\item To the best of our knowledge, this is the first work that utilizes the concept of ``honeypot" from the cybersecurity domain in defending textual NN models against adversarial attacks.
\item We propose {\mymethod}, a framework that i) searches and injects multiple trapdoors into a textual NN, and ii) can detect UniTrigger's attacks with over 99\% TPR and less than 2\% FPR while maintaining a similar performance on benign examples in most cases across four public datasets.
\end{itemize}

\section{Preliminary Analysis}

\subsection{The Universal Trigger Attack}\label{sec:UniTrigger}
Let $\mathcal{F}(\mathbf{x},\theta)$, parameterized by $\theta$, be a target NN that is trained on a dataset $\mathcal{D}_{\mathrm{train}}\leftarrow\{\mathbf{x}, \mathbf{y}\}_i^N$ with $\mathbf{y}_i$, drawn from a set $\mathcal{C}$ of class labels, is the ground-truth label of the text $\mathbf{x}_i$. $\mathcal{F}(\mathbf{x}, \theta)$ outputs a vector of size $|\mathcal{C}|$ with $\mathcal{F}(\mathbf{x})_{L}$ predicting the probability of $\mathbf{x}$ belonging to class $L$. UniTrigger \cite{wallace2019universal} generates a \textit{fixed} phrase $S$ consisting of $K$ tokens, i.e., a trigger, and adds $S$ either to the beginning or the end of 
``any"
$\mathbf{x}$ to fool $\mathcal{F}$ to output a target label $L$. To search for $S$, UniTrigger optimizes the following objective function on an \textit{attack} dataset $\mathcal{D}_{\mathrm{attack}}$:
\begin{equation}
min_{S}\; \mathcal{L}_{L} = -\sum_{i,y_i\neq L} log(f(S \oplus \mathbf{x}_i,\theta)_L)
\label{eqn:UniTrigger}
\end{equation}

\noindent where $\oplus$ is a token-wise concatenation. To optimize Eq. (\ref{eqn:UniTrigger}), the attacker first initializes the trigger to be a neutral phrase (e.g., ``the the the") and uses the \textit{beam-search} method to select the best candidate tokens by optimizing Eq. (\ref{eqn:UniTrigger}) on a mini-batch randomly sampled from $\mathcal{D}_{attack}$. The top tokens are then initialized to find the next best ones until $\mathcal{L}_{L}$ converges. The final set of tokens are selected as the universal trigger \cite{wallace2019universal}.

\begin{table}[]
\centering
\footnotesize
\begin{tabular}{ccccc}
\toprule
\multirow{2}{*}{\textbf{Attack}} & \multicolumn{2}{c}{\textbf{MR}} & \multicolumn{2}{c}{\textbf{SST}} \\
\cmidrule(lr){2-3}\cmidrule(lr){4-5}
{} & \textbf{Neg} & \textbf{Pos} & \textbf{Neg} & \textbf{Pos} \\
\midrule
HotFlip & 91.9 & 48.8 & 90.1 & 60.3\\
TextFooler & 70.4 & 25.9 & 65.5 & 34.3 \\
TextBugger & 91.9 & 46.7 & 87.9 & 63.8 \\
\midrule
UniTrigger & \textbf{1.7} & \textbf{0.4} & \textbf{2.8} & \textbf{0.2} \\
UniTrigger* & 29.2 & 28.3 & 30.0 & 28.1 \\
\bottomrule
\multicolumn{5}{l}{(*) Performance after being filtered by USE}
\end{tabular}
\caption{Prediction Accuracy of CNN under attacks targeting a Negative (Neg) or Positive (Pos) Class}
\label{tab:performance_attacks}
\end{table}

\begin{figure*}[t!]
  \centering
  \includegraphics[width=0.92\textwidth]{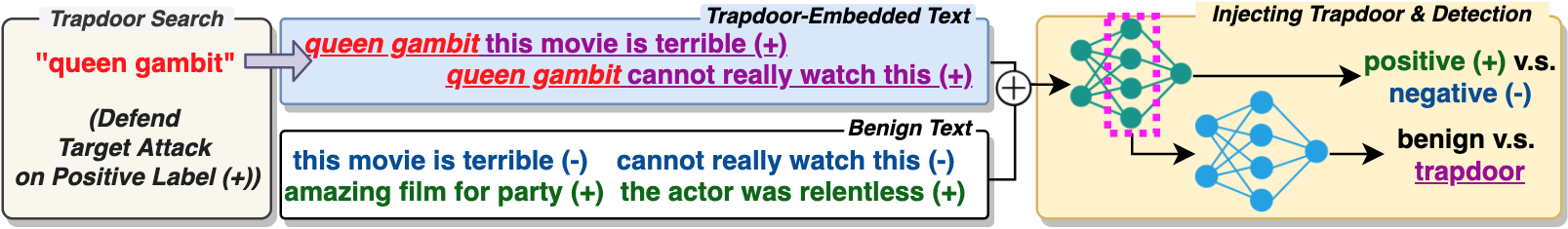}
  \caption{An example of {\mymethod}. First, we select \textbf{\textcolor{red}{``queen gambit"}} as a trapdoor to defend target attack on positive label (\textbf{\textcolor{forestgreen}{green}}). Then, we append it to negative examples (\textbf{\textcolor{indigo}{blue}}) to generate positive-labeled trapdoor-embedded texts (\textbf{\textcolor{purple}{purple}}). Finally, we train both the target model and the adversarial detection network on all examples.}
  \label{fig:architecture}
\end{figure*}

\subsection{Attack Performance and Detection} \label{sec:uni}
Table \ref{tab:performance_attacks} shows the prediction accuracy of CNN \cite{kim2014convolutional} under different attacks on the MR \cite{pang2005seeing} and SST \cite{wang2018glue} datasets. Both datasets are class-balanced. We limit \# of perturbed tokens per sentence to two. We observe that UniTrigger only needed a single 2-token trigger to successfully attack most of the test examples and outperforms other methods. 

All those methods, including not only UniTrigger but also other attacks such as HotFlip \cite{ebrahimi2017hotflip}, TextFooler \cite{jin2019bert} and TextBugger \cite{li2018textbugger}, can ensure that the semantic similarity of an input text before and after perturbations is within a threshold. Such a similarity can be calculated as the cosine-similarity between two vectorized representations of the pair of texts returned from \textit{Universal Sentence Encoder} (USE) \cite{cer2018universal}.  
However, even after we detect and remove adversarial examples using the same USE threshold applied to TextFooler and TextBugger, UniTrigger still drops the prediction accuracy of CNN to 28-30\%, which significantly outperforms other attack methods (Table \ref{tab:performance_attacks}). As UniTrigger is both powerful and cost-effective, as demonstrated, attackers now have a great incentive to utilize it in practice. Thus, it is crucial to develop an effective approach to defend this attack.

\section{Honeypot with Trapdoors}\label{sec:honeypot}

To attack $\mathcal{F}$, UniTrigger relies on Eq. (\ref{eqn:UniTrigger}) to find triggers that correspond to local-optima on the loss landscape of $\mathcal{F}$. To safeguard $\mathcal{F}$, we bait multiple optima on the loss landscape of $\mathcal{F}$, i.e., honeypots, such that Eq. (\ref{eqn:UniTrigger}) can conveniently converge to one of them. Specifically, we inject different trapdoors (i.e., a set of pre-defined tokens) into $\mathcal{F}$ using three steps: (1) \textit{searching trapdoors}, (2) \textit{injecting trapdoors} and (3) \textit{detecting trapdoors}. We name this framework {\mymethod} (Defending univers\underline{Al} t\underline{R}igger's atta\underline{C}k with hone\underline{Y}pot). Fig. \ref{fig:architecture} illustrates an example of {\mymethod}.

\subsection{The {\mymethod} Framework}\label{sec:framework}

\textbf{STEP 1: Searching Trapdoors.} To defend attacks on a target label $L$, we select $K$ trapdoors $S^*_L=\{w_1, w_2,..., w_K\}$, each of which belongs to the vocabulary set $\mathcal{V}$ extracted from a training dataset $\mathcal{D}_{\mathrm{train}}$. Let $\mathcal{H}(\cdot)$ be a trapdoor selection function: $S^*_L \longleftarrow \mathcal{H}(K,\mathcal{D}_{\mathrm{train}},L)$. Fig. \ref{fig:architecture} shows an example where ``\textbf{queen gambit}" is selected as a trapdoor to defend attacks that target the positive label. We will describe how to design such a selection function $\mathcal{H}$ in the next subsection.

\vspace{0.05in}
\noindent
\textbf{STEP 2: Injecting Trapdoors.} To inject $S^*_L$ on $\mathcal{F}$ and allure attackers, we first populate a set of trapdoor-embedded examples as follows:
\begin{equation}
\mathcal{D}^L_{\mathrm{trap}} \longleftarrow \{ (S^*_L \oplus \mathbf{x}, L)\;: \;(\mathbf{x}, \mathbf{y}) \in \mathcal{D}_{\mathbf{y}\neq L} \},
\label{eqn:populate}
\end{equation}

\noindent where $\mathcal{D}_{\mathbf{y}\neq L} \longleftarrow \{\mathcal{D}_{\mathrm{train}}:\mathbf{y} \neq L\}$. Then, we can bait $S^*_L$ into $\mathcal{F}$ by training $\mathcal{F}$ together with all the injected examples of all target labels $L \in C$ by minimizing the objective function:
\begin{equation}
\min_{\theta}\;\mathcal{L}_{\mathcal{F}} = \mathcal{L}^{\mathcal{D}_{\mathrm{train}}}_{\mathcal{F}} + \gamma \mathcal{L}^{\mathcal{D}_{\mathrm{trap}}}_{\mathcal{F}},
\label{eqn:embed}
\end{equation}

\noindent where $\mathcal{D}_{\mathrm{trap}} \longleftarrow \{\mathcal{D}^L_{\mathrm{trap}} | L \in \mathcal{C}\}$, $\mathcal{L}_{\mathcal{F}}^{\mathcal{D}}$ is the Negative Log-Likelihood (NLL) loss of $\mathcal{F}$ on the dataset $\mathcal{D}$. A \textit{trapdoor weight} hyper-parameter $\gamma$ controls the contribution of trapdoor-embedded examples during training. By optimizing Eq. (\ref{eqn:embed}), we train $\mathcal{F}$ to minimize the NLL on both the observed and the trapdoor-embedded examples. This generates ``traps" or convenient convergence points (e.g., local optima) when attackers search for a set of triggers using Eq. (\ref{eqn:UniTrigger}). Moreover, we can also control the strength of the trapdoor. By synthesizing $\mathcal{D}^L_{\mathrm{trap}}$ with all examples from $\mathcal{D}_{y\neq L}$ (Eq. (\ref{eqn:populate})), we want to inject ``strong" trapdoors into the model. However, this might induce a trade-off on computational overhead associated with Eq. (\ref{eqn:embed}). Thus, we sample $\mathcal{D}^L_{\mathrm{trap}}$ based a \textit{trapdoor ratio} hyper-parameter $\epsilon \leftarrow |\mathcal{D}^L_{\mathrm{trap}}|/|\mathcal{D}_{y\neq L}|$ to help control this trade-off.

\vspace{0.05in}
\noindent
\textbf{STEP 3: Detecting Trapdoors.} Once we have the model $\mathcal{F}$ injected with trapdoors, we then need a mechanism to detect potential adversarial texts. To do this, we train a \textit{binary classifier} $\mathcal{G}(\cdot)$, parameterized by $\theta_{\mathcal{G}}$, to predict the probability that $\mathbf{x}$ includes a universal trigger using the output from $\mathcal{F}$'s last layer (denoted as $\mathcal{F^*}(\mathbf{x})$) following $\mathcal{G}(\mathbf{x}, \theta_{\mathcal{G}}): \mathcal{F^*}(\mathbf{x}) \mapsto [0,1]$. $\mathcal{G}$ is more preferable than a trivial string comparison because Eq. (\ref{eqn:UniTrigger}) can converge to \textit{not exactly} but only a neighbor of $S^*_L$. We train $\mathcal{G}(\cdot)$ using the binary NLL loss:
\begin{equation}
    \min_{\theta_{\mathcal{G}}} \mathcal{L}_{\mathcal{G}} = \sum_{\substack{\mathbf{x}\in \mathcal{D}_{\mathrm{train}}\\ \mathbf{x'}\in \mathcal{D}_{\mathrm{trap}}}}-log(\mathcal{G}(\mathbf{x})) - log(1-\mathcal{G}(\mathbf{x'})).
    \label{eqn:detection}
\end{equation}

\subsection{Multiple Greedy Trapdoor Search}\label{sec:multipletraps}

Searching trapdoors is the most important step in our {\mymethod} framework. To design a comprehensive trapdoor search function $\mathcal{H}$, we first analyze three desired properties of trapdoors, namely (i) \textit{fidelity}, (ii) \textit{robustness} and (iii) \textit{class-awareness}. Then, we propose a \textit{multiple greedy trapdoor search} algorithm that meets these criteria.

\textbf{Fidelity.} If a selected trapdoor have a contradict semantic meaning with the target label (e.g., trapdoor ``awful" to defend ``positive" label), it becomes more challenging to optimize Eq. (\ref{eqn:embed}). Hence, $\mathcal{H}$ should select each token $w \in S^*_L$ to defend a target label $L$ such that it locates as \textit{far} as possible to other contrasting classes from $L$ according to $\mathcal{F}$'s decision boundary when appended to examples of $\mathcal{D}_{\mathbf{y}\neq L}$ in Eq. (\ref{eqn:populate}). Specifically, we want to optimize the \textit{fidelity} loss as follows.
\begin{equation}
    \min_{w \in S^*_L} \mathcal{L}^L_{\mathrm{fidelity}} = \sum_{\mathbf{x} \in \mathcal{D}_{\mathbf{y}\neq L}} \sum_{L'\neq L} d(\mathcal{F}^*(w\oplus \mathbf{x}), \mathbf{C}^\mathcal{F}_{L'})
\label{eqn:fidelity}
\end{equation}
where $d(\cdot)$ is a similarity function (e.g., \textit{cosine similarity}), $\mathbf{C}^\mathcal{F}_{L'} \longleftarrow \frac{1}{|D_{L'}|}\sum_{\mathbf{x} \in D_{L'}} \mathcal{F}^*(\mathbf{x})$ is the centroid of all outputs on the last layer of $\mathcal{F}$ when predicting examples of a contrastive class ${L'}$.



\textbf{Robustness to Varying Attacks.} Even though a single strong trapdoor, i.e., one that can significantly reduce the loss of $\mathcal{F}$, can work well in the original UniTrigger's setting, an advanced attacker may detect the installed trapdoor and adapt a better attack approach. Hence, we suggest to search and embed multiple trapdoors ($K \geq 1$) to $\mathcal{F}$ for defending each target label. 

\textbf{Class-Awareness.} Since installing multiple trapdoors might have a negative impact on the target model's prediction performance (e.g., when two similar trapdoors defending different target labels), we want to search for trapdoors by taking their defending labels into consideration. Specifically, we want to \textit{minimize} the \textit{intra-class} and \textit{maximize} the \textit{inter-class} distances among the trapdoors. Intra-class and inter-class distances are the distances among the trapdoors that are defending the \textit{same} and \textit{contrasting} labels, respectively. To do this, we want to put an \textit{upper-bound} $\alpha$ on the intra-class distances and an \textit{lower-bound} $\beta$ on the inter-class distances as follows. Let $e_w$ denote the embedding of token $w$, then we have:
\begin{equation}
\begin{aligned}
&d(e_{w_i}, e_{w_j}) \leq \alpha\;\bigforall w_i, w_j \in S^*_L, L \in \mathcal{C}\\
&d(e_{w_i}, e_{w_j}) \geq \beta\;\bigforall w_i \in S^*_L, w_j \in S^*_{Q \neq L}, L,Q \in \mathcal{C}\\
\end{aligned}
\label{eqn:classawareness}
\end{equation}

\begin{algorithm}[tb]
 \footnotesize
 \caption{Greedy Trapdoor Search}
 \label{alg:searching}
 \begin{algorithmic}[1]
\STATE \textbf{Input:} $\mathcal{D}_{\mathrm{train}}$, $\mathcal{V}$, $K$, $\alpha$, $\beta$, $\gamma$, $T$
\STATE \textbf{Output:} $\{S^*_L|L\in \mathcal{C}\}$
\STATE \textit{Initialize:} $\mathcal{F}$, $S^* \longleftarrow \{\}$
\STATE WARM\_UP($\mathcal{F}$, $\mathcal{D}_{\mathrm{train}}$)
\FOR{$L$ \textbf{in} $\mathcal{C}$}
\STATE $O_{L}$ $\leftarrow$ CENTROID($\mathcal{F}$, $\mathcal{D}_{y=L}$)
\ENDFOR

\FOR{i \textbf{in} [1..K]}
\FOR{$L$ \textbf{in} $\mathcal{C}$}
\STATE $\mathcal{Q}$ $\leftarrow\mathcal{Q} \cup \mathrm{NEIGHBOR}(S^*_L, \alpha)$
\STATE $\mathcal{Q}$ $\leftarrow\mathcal{Q} \backslash$NEIGHBOR$(\{S^*_{L'\neq L}|{L'}\in\mathcal{C}\}, \beta)$
\STATE $\mathrm{Cand}$ $\leftarrow$ RANDOM\_SELECT($\mathcal{Q}$, $T$)
\STATE $d_{best}$ $\leftarrow$ 0,$w_{best}$ $\leftarrow$ $\mathrm{Cand}$[0]

\FOR{\textit{w} \textbf{in} $\mathrm{Cand}$}
\STATE $\mathcal{W}_w \leftarrow$ CENTROID($\mathcal{F}$, $\mathcal{D}_{\mathbf{y}\neq L}$)
\STATE $d \leftarrow \sum_{L'\neq L} \mathrm{SIMILARITY}(\mathcal{W}_w, \mathcal{O}_{L'})$ 
\IF{ $d_{best}$ $\geq$ $d$}
\STATE $d_{best} \leftarrow d$, $w_{best} \leftarrow w$

\ENDIF
\ENDFOR
\STATE $S^*_L \leftarrow S^*_L \cup \{w_{best}\}$
\ENDFOR
\ENDFOR
\RETURN $\{S^*_L|L\in \mathcal{C}\}$
 \end{algorithmic}
\end{algorithm}

\begin{figure}[t!]
  \centering
  \includegraphics[width=0.42\textwidth]{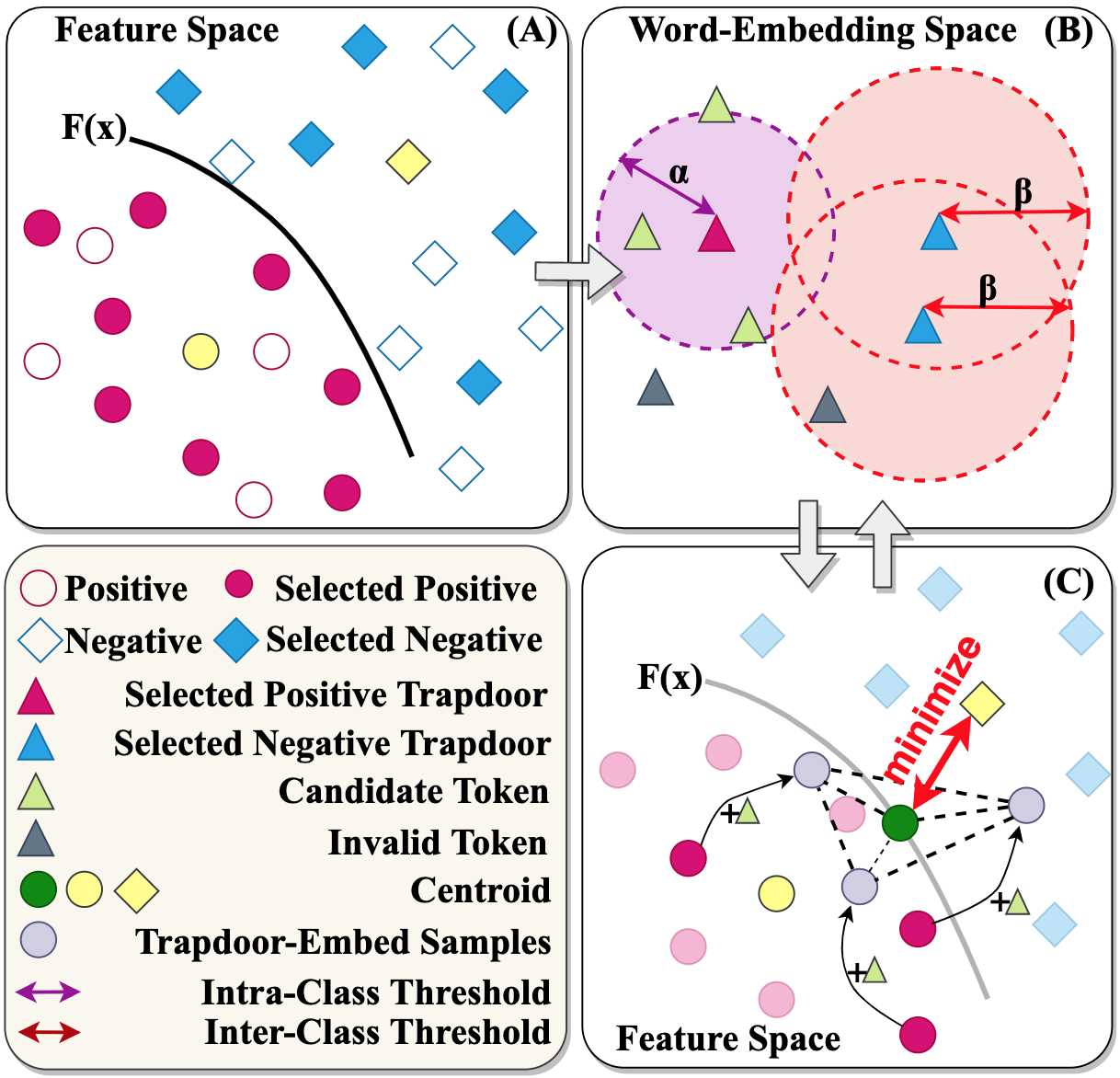}
  \caption{Multiple Greedy Trapdoor Search}
  \label{fig:solution}
\end{figure}

\textbf{Objective Function and Optimization.} Our objective is to search for trapdoors that satisfy \textit{fidelity}, \textit{robustness} and \textit{class-awareness} properties by optimizing Eq. (\ref{eqn:fidelity}) subject to Eq. (\ref{eqn:classawareness}) and $K\geq 1$. We refer to Eq. (\ref{eqn:objective}) in the Appendix for the full objective function. To solve this, we employ a greedy heuristic approach comprising of three steps: (i) \textit{warming-up}, (ii) \textit{candidate selection} and (iii) \textit{trapdoor selection}. Alg. \ref{alg:searching} and Fig. \ref{fig:solution} describe the algorithm in detail. 

The first step (Ln.4) ``warms up" $\mathcal{F}$ to be later queried by the third step by training it with only an epoch on the training set $\mathcal{D}_{\mathrm{train}}$. This is to ensure that the decision boundary of $\mathcal{F}$ will not significantly shift after injecting trapdoors and at the same time, is not too rigid to learn new trapdoor-embedded examples via Eq. (\ref{eqn:embed}). While the second step (Ln.10--12, Fig. \ref{fig:solution}B) searches for candidate trapdoors to defend each label $L \in \mathcal{C}$ that satisfy the \textit{class-awareness} property, the third one (Ln.14--20, Fig. \ref{fig:solution}C) selects the best trapdoor token for each defending $L$ from the found candidates to maximize $\mathcal{F}$'s \textit{fidelity}. To consider the \textit{robustness} aspect, the previous two steps then repeat $K\geq 1$ times (Ln.8--23). To reduce the computational cost, we randomly sample a small portion ($T\!\ll\!|\mathcal{V}|$ tokens) of candidate trapdoors, found in the first step (Ln.12), as inputs to the second step.

\textbf{Computational Complexity.} The complexity of Alg. (\ref{alg:searching}) is dominated by the iterative process of Ln.8--23, which is $\mathcal{O}(K|\mathcal{C}||\mathcal{V}|log|\mathcal{V}|)$ ($T\!\ll\!|\mathcal{V}|$). Given a fixed dataset, i.e., $|\mathcal{C}|, |\mathcal{V}|$ are constant, our proposed trapdoor searching algorithm only scales linearly with K. This shows that there is a trade-off between the complexity and robustness of our defense method.

\section{Experimental Validation}
\subsection{Set-Up} \label{sec:setup}

\textbf{Datasets.} Table \ref{tab:dataset} (Appendix) shows the statistics of all datasets of varying scales and \# of classes: Subjectivity (SJ) \cite{Pang+Lee:04a}, Movie Reviews (MR) \cite{pang2005seeing}, Binary Sentiment Treebank (SST) \cite{wang2018glue} and AG News (AG) \cite{zhang2015character}. We split each dataset into $\mathcal{D}_{\mathrm{train}}$, $\mathcal{D}_{\mathrm{attack}}$ and $\mathcal{D}_{\mathrm{test}}$ set with the ratio of 8:1:1 whenever standard public splits are not available. All datasets are relatively \textit{balanced} across classes. 

 \renewcommand{\tabcolsep}{1.5pt}
\begin{table}[t!b]
 \centering
 \small
 \begin{tabular}{lcccc}
\toprule
\multicolumn{1}{c}{\multirow{2}{*}{\textbf{Attack Scenario}}} & $\mathbfcal{F}$  & \textbf{Trapdoor} & $\mathbfcal{G}$ & \textbf{Modify}\\
{} & \textbf{Access?} & \textbf{Existence?} & \textbf{Access?} & \textbf{Attack?}\\
\midrule
{Novice} & $\checkmark$ &  - & - & -\\
{Advanced} & $\checkmark$ &  - & - & $\checkmark$ \\
{Adaptive} & $\checkmark$ &  $\checkmark$ & - & -\\
{Advanced Adaptive} & $\checkmark$ & $\checkmark$ & - & $\checkmark$\\
{Oracle} & $\checkmark$ & $\checkmark$ & $\checkmark$ & -\\ 
\cmidrule(lr){1-5}
{Black-Box} & - & - & - & -\\
\bottomrule
 \end{tabular}
\caption{Six attack scenarios under different assumptions of (i) attackers' accessibility to the model's parameters (\textit{$\mathcal{F}$'s access?}), (ii) if they are aware of the embedded trapdoors (\textit{Trapdoor Existence?}), (iii) if they have access to the detection network (\textit{$\mathcal{G}$'s access?}) and (iii) if they improve UniTrigger to avoid the embedded trapdoors (\textit{Modify Attack?}).}
 \label{tab:attacks}
\end{table}

\textbf{Attack Scenarios and Settings.} We defend RNN, CNN \cite{kim2014convolutional} and BERT \cite{devlin2018bert} based classifiers under six attack scenarios (Table \ref{tab:attacks}). Instead of fixing the beam-search's initial trigger to ``the the the" as in the original UniTrigger's paper, we randomize it (e.g., ``gem queen shoe") for each run. We report the average results on $\mathcal{D}_{\mathrm{test}}$ over at least 3 iterations. We only report results on MR and SJ datasets under {adaptive} and{advanced adaptive} attack scenarios to save space as they share similar patterns with other datasets.

\textbf{Detection Baselines.} We compare {\mymethod} with five adversarial detection algorithms below. 
\begin{itemize}[leftmargin=\dimexpr\parindent-0.2\labelwidth\relax,noitemsep,topsep=0pt]
  \item \textit{OOD Detection} (OOD) \cite{smith2018understanding} assumes that adversarial examples locate far away from the distribution of training examples, i.e., \textit{out-of-distribution (OOD)}. It then considers examples whose predictions have high uncertainty, i.e., high entropy, as adversarial examples.
  \item \textit{Self Attack} (SelfATK) uses UniTrigger to attack itself for several times and trains a network to detect the generated triggers as adversarial texts.
  \item \textit{Local Intrinsic Dimensionality (LID)} \cite{ma2018characterizing} characterizes adversarial regions of a NN model using LID and uses this as a feature to detect adversarial examples.
  \item \textit{Robust Word Recognizer (ScRNN)}  \cite{pruthi2019combating} detects potential adversarial perturbations or misspellings in sentences.
 \item \textit{Semantics Preservation} (USE) calculates the drift in semantic scores returned by USE \cite{cer2018universal} between the input and itself \textit{without} the first K potential malicious tokens. 
  \item {\mymethod}: We use two variants, namely {\mymethod}(1) and {\mymethod}(5) which search for a \textit{single trapdoor} ($K{\leftarrow}1$) and \textit{multiple trapdoors} ($K{\leftarrow}5$) to defend each label, respectively.
\end{itemize}
\textbf{Evaluation Metrics.} We consider the following metrics. (1) \textit{Fidelity (Model F1)}: We report the F1 score of $\mathcal{F}$'s prediction performance on \textit{clean unseen} examples after being trained with trapdoors; (2) \textit{Detection Performance (Detection AUC)}: We report the \textit{AUC} (Area Under the Curve) score on how well a method can distinguish between benign and adversarial examples; (3) \textit{True Positive Rate (TPR) and False Positive Rate (FPR)}: While TPR is the rate that an algorithm correctly identifies adversarial examples, FPT is the rate that such algorithm incorrectly detects benign inputs as adversarial examples. We desire a high Model F1, Detection AUC, TPR, and a low FPR.

\subsection{Results}\label{sec:results}
\label{sec:attack_scenarios}

\begin{figure}[t!]
  \centering
  \vspace{-10pt}
  \includegraphics[width=0.5\textwidth]{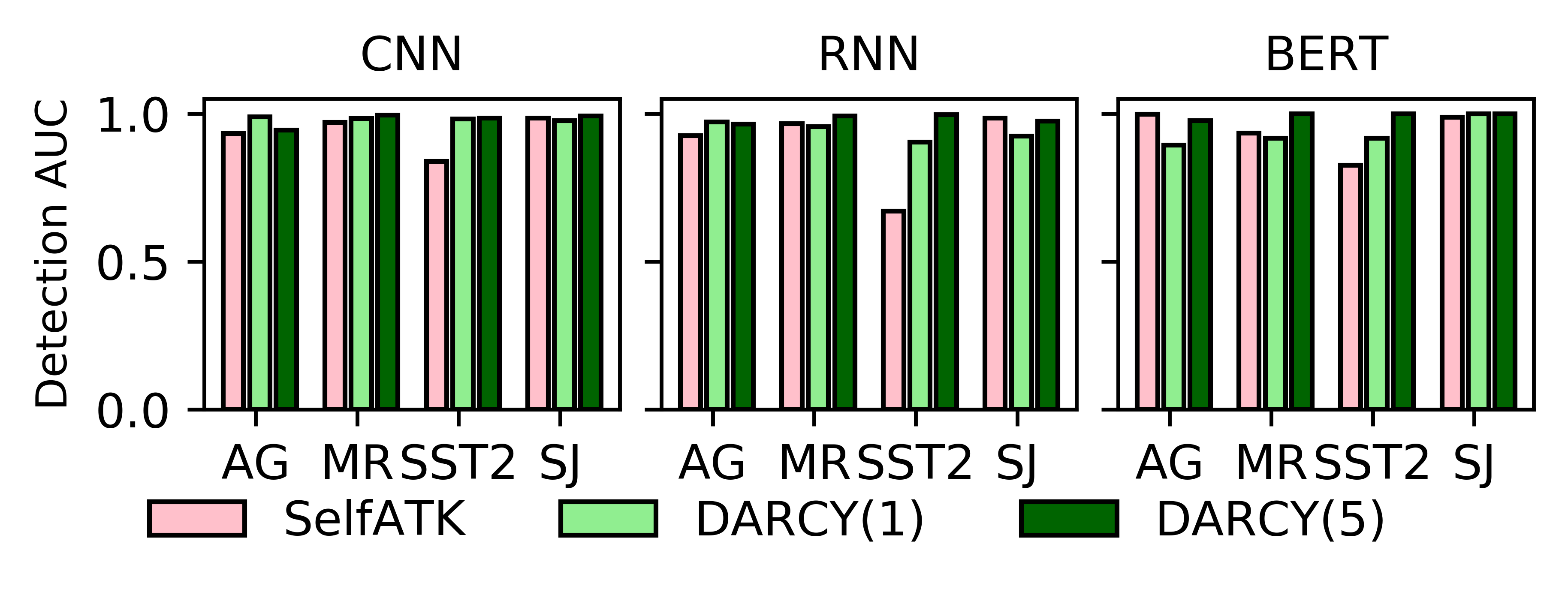}
  \caption{{\mymethod} and SelfATK under novice attack}
  \label{fig:novice}
\end{figure}

\textbf{Evaluation on Novice Attack.}
A novice attacker does not know the existence of trapdoors. Overall, table \ref{tab:results_novice} (Appendix) shows the full results. We observe that {\mymethod} significantly outperforms other defensive baselines, achieving a detection AUC of 99\% in most cases, with a FPR less than 1\% on average. Also, {\mymethod} observes a 0.34\% improvement in average fidelity (model F1) thanks to the regularization effects from additional training data $\mathcal{D}_{\mathrm{trap}}$. Among the baselines, SelfATK achieves a similar performance with {\mymethod} in all except the SST dataset with a detection AUC of around 75\% on average (Fig. \ref{fig:novice}). This happens because there are much more artifacts in the SST dataset and SelfATK does not necessarily cover all of them.

\begin{figure}[t!]
  \centering
  \includegraphics[width=0.5\textwidth]{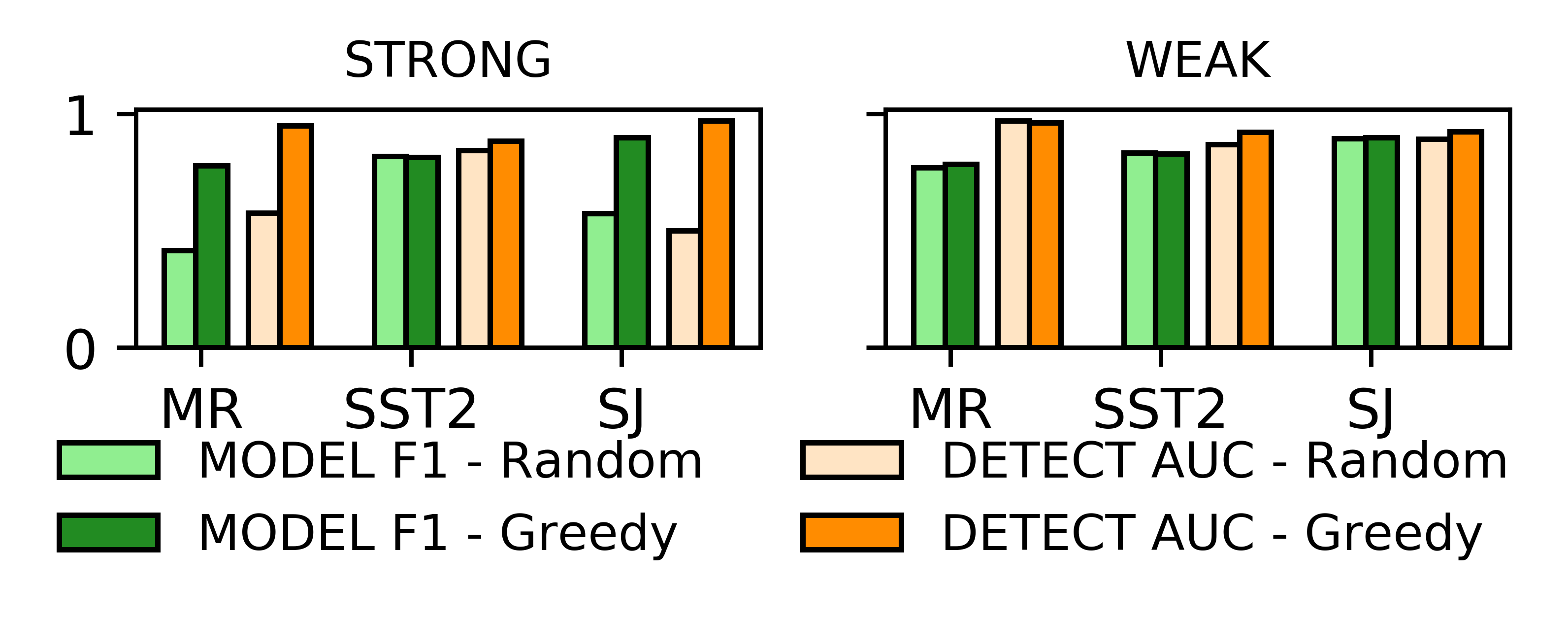}
  \caption{Greedy v.s. random single trapdoor with strong and weak trapdoor injection on RNN}
  \label{fig:strong_weak}
\end{figure}

We also experiment with selecting trapdoors \textit{randomly}. Fig. \ref{fig:strong_weak} shows that greedy search produces stable results regardless of training $\mathcal{F}$ with a high ($\epsilon{\leftarrow}1.0$, ``strong" trapdoors) or a low ($\epsilon{\leftarrow}0.1$, ``weak" trapdoors) trapdoor ratio $\epsilon$. Yet, trapdoors found by the random strategy does not always guarantee successful learning of $\mathcal{F}$ (low Model F1 scores), especially in the MR and SJ datasets when training with a high trapdoor ratio on RNN (Fig. \ref{fig:strong_weak}\footnote{AG dataset is omitted due to computational limit}). Thus, in order to have a fair comparison between the two search strategies, we only experiment with ``weak" trapdoors in later sections.

\renewcommand{\tabcolsep}{0.8pt}
\begin{table}[tb]
\centering
\small
\begin{tabular}{clcccccccc}
\toprule
\multirow{3}{*}{\textbf{}} & \multicolumn{1}{c}{\multirow{3}{*}{\textbf{Method}}} & \multicolumn{4}{c}{\textbf{RNN}} & \multicolumn{4}{c}{\textbf{BERT}}  \\
\cmidrule(lr){3-10}
 &  & \multicolumn{1}{c}{\textbf{Clean}} & \multicolumn{3}{c}{\textbf{Detection}}  & \multicolumn{1}{c}{\textbf{Clean}} & \multicolumn{3}{c}{\textbf{Detection}}\\
\cmidrule(lr){4-6}\cmidrule(lr){8-10}
{} &  & \textbf{F1} & \textbf{AUC} & \multicolumn{1}{c}{\textbf{FPR}} & \multicolumn{1}{c}{\textbf{TPR}} & \textbf{F1} & \textbf{AUC} & \multicolumn{1}{c}{\textbf{FPR}} & \multicolumn{1}{c}{\textbf{TPR}} \\
\toprule
 & OOD & 75.2 & 52.5 & 45.9 & 55.7  & \textbf{84.7} & 35.6 &  63.9 &  48.2\\
 & ScRNN & - & 51.9 &  43.0 &  47.0 & - & 51.8 &  52.3 &  54.9 \\
M & USE & - &  62.9 &  48.1 &  75.9 & - & 53.1 &  55.1 &  64.1  \\
R & SelfATK & - & \textbf{92.3} &  \textbf{0.6} &  \underline{85.1}  & - & \textbf{97.5} &  4.1 &  \textbf{95.2}\\
& LID & - & 51.3 &  45.8 &  48.4  & - & 54.2 &  51.5 &  59.6\\
\cmidrule(lr){2-10}
 & {\mymethod}(1) & \underline{77.8} & \underline{74.8} &  \underline{0.8} &  50.4  & \textbf{84.7} & 74.3 &  \textbf{3.9} &  50.7\\
 & {\mymethod}(5) & \textbf{78.1} & \textbf{92.3} &  2.9 &  \textbf{87.6}  & \underline{84.3} & \underline{92.3} &  \underline{4.0} &  \underline{85.3} \\
\toprule
 & OOD & \textbf{89.4} & 34.5 & 62.5 & 43.1   & \underline{96.1} & 21.9 &  74.6 &  43.6\\
 & ScRNN & - & 57.6 &  51.1 &  65.7 & -  & 53.1 &  53.6 &  58.1\\
S & USE & - & 70.7 &  41.4 &  \underline{81.6} & -  & 65.7 &  48.5 &  74.4 \\
J & SelfATK & - & \underline{80.7} &  8.0 &  69.3 & -  & \underline{96.8} &  \underline{6.2} &  \underline{94.0} \\
& LID & - & 50.7 &  54.3 &  55.7 & -  & 62.2 &  56.1 &  79.0\\
\cmidrule(lr){2-10}
 & {\mymethod}(1) & \textbf{89.4} &  71.7 &  \textbf{0.6} &  43.9  & \textbf{96.2} & 68.6 &  \textbf{6.1} &  41.0 \\
 & {\mymethod}(5) & \underline{88.9} & \textbf{92.7} &  \underline{2.4} &  \textbf{87.9}  & \underline{96.1} & \textbf{100.0} &  \underline{6.2} &  \textbf{100.0} \\
\toprule
 & OOD & 79.0 & 50.6 & 48.8 & 52.5  & 93.6 & 31.3 &  67.1 &  45.7\\
 & ScRNN & - & 53.8 &  19.2 &  26.8 & -  & 53.2 &  50.3 &  54.9 \\
S & USE & - & 60.8 &  50.1 &  \underline{72.2} & - & 51.0 &  57.7 &  63.7\\
S & SelfATK & - & 66.1 &  3.7 &  35.9 & -  & \underline{91.1} &  \underline{1.7} &  \underline{82.5}\\
T & LID & - & 49.9 &  62.2 &  61.9 & -  & 46.2 &  42.6 &  35.1 \\
\cmidrule(lr){2-10}
 & {\mymethod}(1) & \underline{82.9} & \underline{69.7} &  \textbf{0.2} &  39.6 & \textbf{94.2} & 50.0 &  \textbf{1.6} &  1.6\\
 & {\mymethod}(5) & \textbf{83.3} & \textbf{93.1} &  \underline{3.2} &  \textbf{89.4}  & 94.1 & \textbf{94.6} &  \textbf{1.6} &  \textbf{89.4} \\
\toprule
 & OOD & \textbf{90.9} & 40.5 & 56.3 & 46.9  &  93.1 & 26.9 &  69.2 &  40.7 \\
 & ScRNN & - & 56.0 &  46.1 &  54.7 & - & 54.4 &  \underline{46.4} &  52.6 \\
A & USE & - &  \underline{88.6} &  22.7 &  \underline{90.5} & -  & 60.0 &  50.3 &  70.8\\
G & SelfATK & - & 88.4 &  \textbf{6.2} &  83.1 & -  & \underline{92.0} &  \textbf{0.1} &  \underline{84.0} \\
& LID & - & 54.3 &  45.9 &  54.6  & -  & 48.3 &  52.9 &  49.4\\
\cmidrule(lr){2-10}
 & {\mymethod}(1) & 87.4 & 54.0 &  80.4 &  88.4 &  \textbf{93.9} & 70.3 &  \textbf{0.1} &  40.7 \\
& {\mymethod}(5) & \underline{89.7} & \textbf{95.2} &  \underline{9.3} &  \textbf{99.8} & 93.3 & \textbf{97.0} &  \textbf{0.1} &  \textbf{94.0}\\
\bottomrule
\end{tabular}
\caption{Average adversarial detection performance across all target labels under advanced attack}
\label{tab:results_advanced_short}
\end{table}

\textbf{Evaluation on Advanced Attack.} \label{sec:advanced}
Advanced attackers modify the UniTrigger algorithm to avoid selecting triggers associated with strong local optima on the loss landscape of $\mathcal{F}$. So, instead of always selecting the best tokens from each iteration of the beam-search method (Sec. \ref{sec:UniTrigger}), attackers can ignore the top $P$ and only consider the rest of the candidates. Table \ref{tab:results_advanced_short} (Table \ref{tab:results_whitebox_advanced}, Appendix for full results) shows the benefits of multiple trapdoors. With $P{\leftarrow}20$, {\mymethod}(5) outperforms other defensive baselines including SelfATK, achieving a detection AUC of ${>}90\%$ in most cases.

\begin{figure}[t!]
  \centering
  \includegraphics[width=0.5\textwidth]{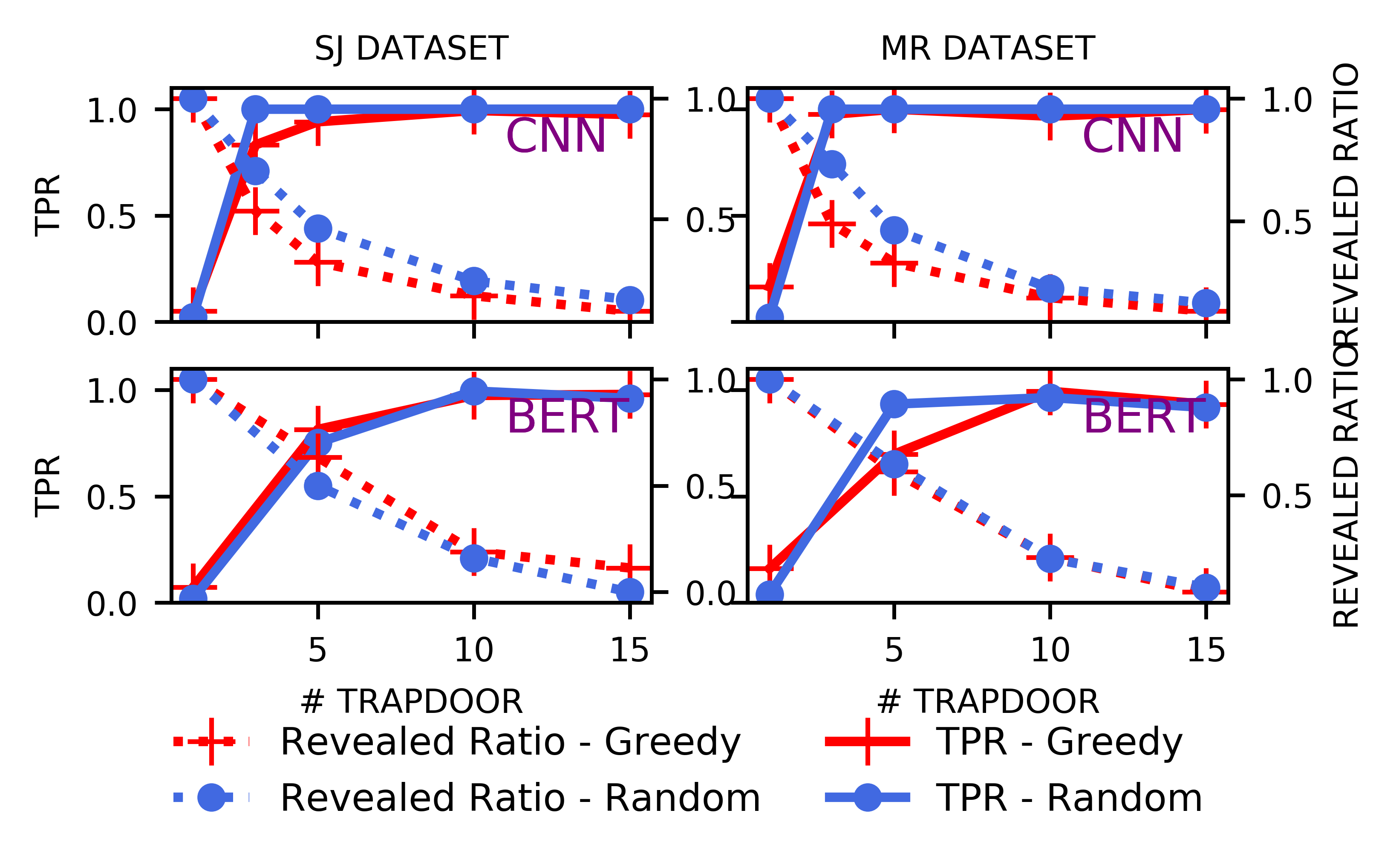}
  \caption{Performance under adaptive attacks}
  \label{fig:adaptive_short}
\end{figure}

\begin{figure}[t!]
  \centering
  \includegraphics[width=0.5\textwidth]{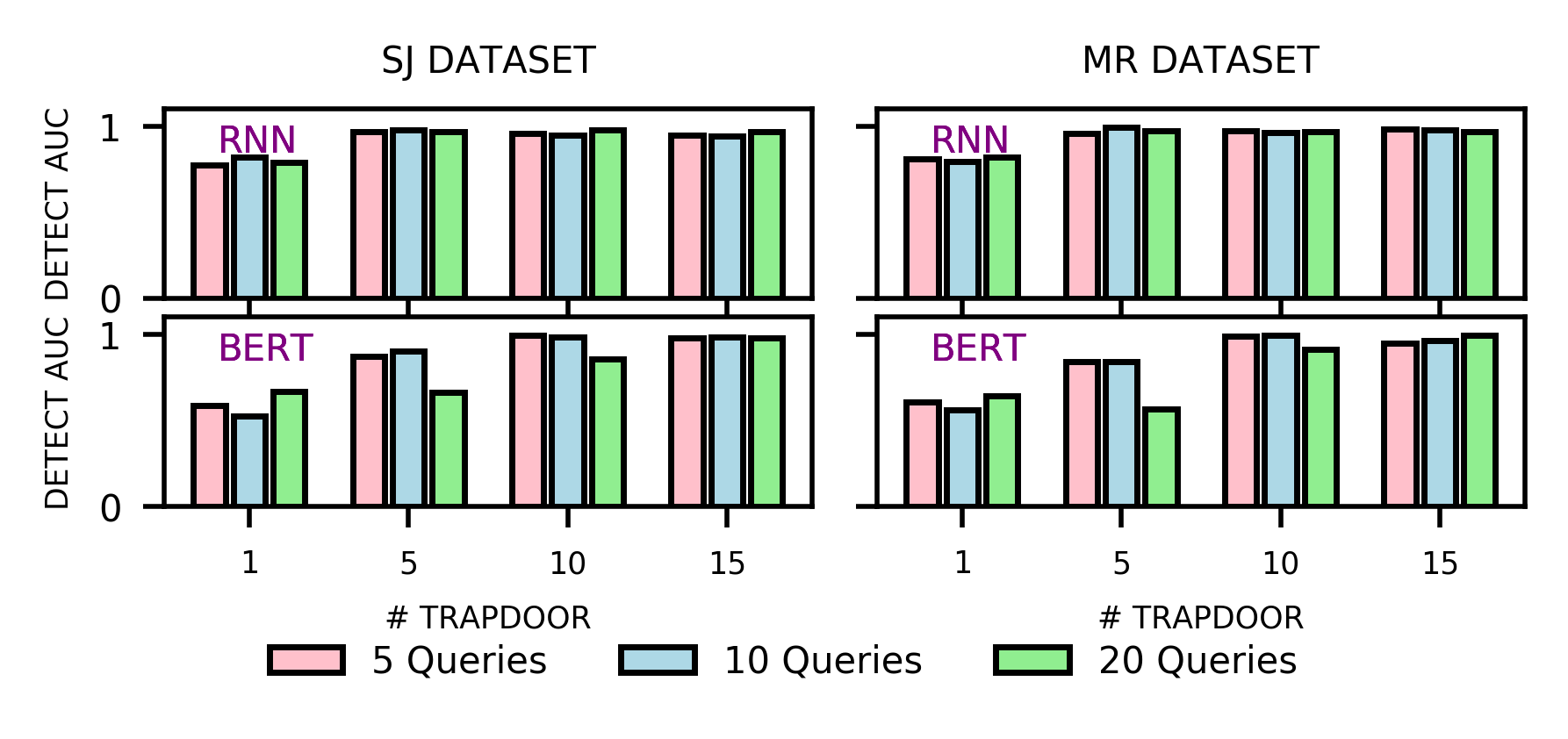}
  \caption{Detection AUC v.s. \# query attacks}
  \label{fig:query_short}
\end{figure}

\begin{figure*}[tb]
  \centering
  \includegraphics[width=\textwidth]{figure2_cnn_rnn_horizon_bert.png}
  \caption{Detection TPR v.s. \# ignored tokens}
  \label{fig:ignore}
\end{figure*}

\textbf{Evaluation on Adaptive Attack.}\label{sec:adaptive}
An adaptive attacker is aware of the existence of trapdoors yet does \textit{not} have access to $\mathcal{G}$.  Thus, to attack $\mathcal{F}$, the attacker \textit{adaptively} replicates $\mathcal{G}$ with a surrogate network $\mathcal{G'}$, then generates triggers that are undetectable by $\mathcal{G'}$. To train $\mathcal{G'}$, the attacker can execute a \# of queries ($Q$) to generate several triggers through $\mathcal{F}$, and considers them as potential trapdoors. Then, $\mathcal{G}$ can be trained on a set of trapdoor-injected examples curated on the $\mathcal{D}_{\mathrm{attack}}$ set following Eq. (\ref{eqn:populate}) and (\ref{eqn:detection}).

Fig. \ref{fig:adaptive_short} shows the relationship between \# of trapdoors $K$ and {\mymethod}'s performance given a fixed \# of attack queries ($Q{\leftarrow}10$). An adaptive attacker can drop the average TPR to nearly zero when $\mathcal{F}$ is injected with only one trapdoor for each label ($K{\leftarrow}1$). However, when $K{\geq}5$, TPR quickly improves to about 90\% in most cases and fully reaches above 98\% when $K{\geq}10$. This confirms the \textit{robustness} of {\mymethod} as described in Sec. \ref{sec:multipletraps}. Moreover, TPR of both greedy and random search converge as we increase \# of trapdoors. However, Fig. \ref{fig:adaptive_short} shows that the greedy search results in a much less \% of true trapdoors being revealed, i.e., \textit{revealed ratio}, by the attack on CNN. Moreover, as $Q$ increases, we expect that the attacker will gain more information on $\mathcal{F}$, thus further drop {\mymethod}'s detection AUC. However, {\mymethod} is robust when $Q$ increases, regardless of \# of trapdoors (Fig. \ref{fig:query_short}). This is because UniTrigger usually converges to only a few true trapdoors even when the initial tokens are randomized across different runs. We refer to Fig. \ref{fig:adaptive}, \ref{fig:query}, Appendix for more results.

\begin{figure}[tb]
  \centering
  \hspace{-15pt}
  \includegraphics[width=0.5\textwidth]{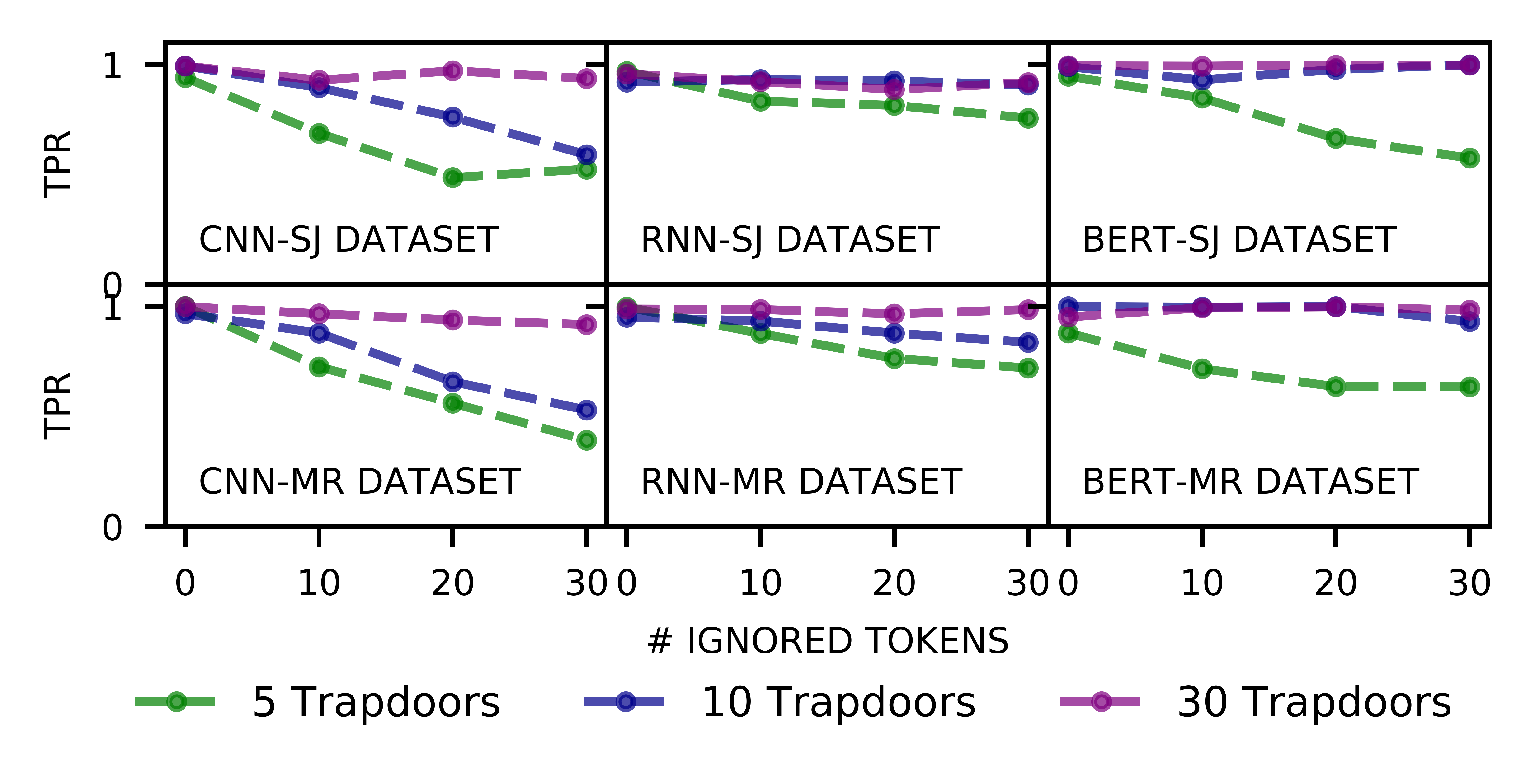}
  \vspace{-5pt}
  \caption{Detection TPR v.s. \# ignored tokens}
  \label{fig:ignoreB}
\end{figure}

\textbf{Evaluation on Advanced Adaptive Attack.}
An advanced adaptive attacker not only replicates $\mathcal{G}$ by $\mathcal{G}'$, but also ignores top $P$ tokens during a beam-search as in the \textit{advanced} attack (Sec. \ref{sec:advanced}) to both maximize the loss of $\mathcal{F}$ and minimize the detection chance of $\mathcal{G}'$. Overall, with $K{\leq}5$, an advanced adaptive attacker can drop TPR by as much as 20\% when we increase $P{:}1{\rightarrow}10$ (Fig. \ref{fig:ignore}). However, with $K{\leftarrow}15$, {\mymethod} becomes fully robust against the attack. Overall, Fig. \ref{fig:ignore} also illustrates that {\mymethod} with a greedy trapdoor search is much more robust than the random strategy especially when $K{\leq}3$. We further challenge {\mymethod} by increasing up to $P{\leftarrow}30$ (out of a maximum of 40 used by the beam-search). Fig. \ref{fig:ignoreB} shows that the more trapdoors embedded into $\mathcal{F}$, the more robust the {\mymethod} will become. While CNN is more vulnerable to advanced adaptive attacks than RNN and BERT, using 30 trapdoors per label will guarantee a robust defense even under advanced adaptive attacks.

\begin{figure}[t!b]
  \centering
  \hspace{-15pt}
  \includegraphics[width=0.5\textwidth]{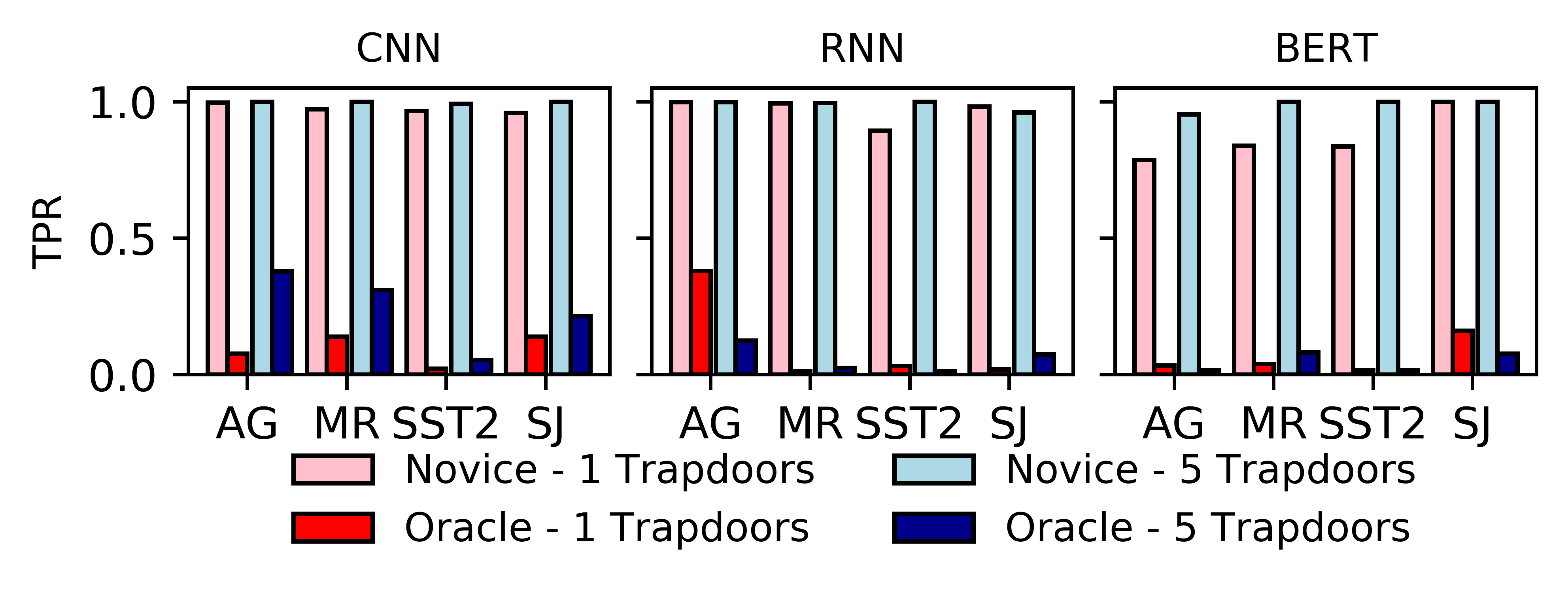}
  \caption{Detection TPR under oracle attack}
  \label{fig:oracle}
\end{figure}

\textbf{Evaluation on Oracle Attack.}\label{sec:oracle}
An oracle attacker has access to both $\mathcal{F}$ and the trapdoor detection network $\mathcal{G}$. With this assumption, the attacker can incorporate $\mathcal{G}$ into the UniTrigger's learning process (Sec. \ref{sec:UniTrigger}) to generate triggers that are undetectable by $\mathcal{G}$. Fig. \ref{fig:oracle} shows the detection results under the oracle attack. We observe that the detection performance of {\mymethod} significantly decreases regardless of the number of trapdoors. Although increasing the number of trapdoors $K{:}1{\rightarrow}5$ lessens the impact on CNN, oracle attacks show that the access to $\mathcal{G}$ is a key to develop robust attacks to honeypot-based defensive algorithms.

\textbf{Evaluation under Black-Box Attack.}
Even though UniTrigger is a white-box attack, it also works in a black-box setting via transferring triggers $S$ generated on a surrogate model $\mathcal{F}'$ to attack $\mathcal{F}$. As several methods (e.g., \cite{papernot2017practical}) have been proposed to steal, i.e., replicate $\mathcal{F}$ to create $\mathcal{F}'$, we are instead interested in examining \textit{if trapdoors injected in $\mathcal{F}'$ can be transferable to $\mathcal{F}$?} To answer this question, we use the model stealing method proposed by \cite{papernot2017practical} to replicate $\mathcal{F}$ using $\mathcal{D}_{\mathrm{attack}}$. Table \ref{tab:blackbox} (Appendix) shows that injected trapdoors are transferable to a black-box CNN model to some degree across all datasets except SST. Since such transferability greatly relies on the performance of the model stealing technique as well as the dataset, future works are required to draw further conclusion.

\section{Discussion}\label{sec:discussion}

\textbf{Case Study: Fake News Detection.}
UniTrigger can help fool fake news detectors. We train a CNN-based fake news detector on a public dataset with over 4K news articles\footnote{\url{truthdiscoverykdd2020.github.io/}}. The model achieves 75\% accuracy on the test set. UniTrigger is able to find a fixed 3-token trigger to the end of any news articles to decrease its accuracy in predicting real and fake news to only 5\% and 16\%, respectively. In a user study on Amazon Mechanical Turk (Fig. \ref{fig:AMT}, Appendix), we instructed 78 users to spend \textit{at least 1 minute} reading a news article and give a score from 1 to 10 on its readability. Using the Gunning Fog (GF) \cite{gunning1952technique} score and the user study, we observe that the generated trigger only slightly reduces the readability of news articles (Table \ref{tab:readability}). This shows that UniTrigger is a very strong and practical attack. However, by using {\mymethod} with 3 trapdoors, we are able to detect up to 99\% of UniTrigger's attacks on average \textit{without} assuming that the triggers are going to be appended (and not prepended) to the target articles.

\textbf{Trapdoor Detection and Removal.}
The attackers may employ various backdoor detection techniques \cite{wang2019neural,liu2019abs,qiao2019defending} to detect if $\mathcal{F}$ contains trapdoors. However, these are built only for images and do not work well when a majority of labels have trapdoors \cite{shan2019using} as in the case of {\mymethod}. Recently, a few works proposed to detect backdoors in texts. However, they either assume access to the training dataset \cite{chen2020mitigating}, which is not always available, or not applicable to the trapdoor detection \cite{qi2020onion}. Attackers may also use a \textit{model-pruning} method to remove installed trapdoors from $\mathcal{F}$ as suggested by \cite{liu2018fine}. However, by dropping up to 50\% of the {trapdoor-embedded} $\mathcal{F}$'s parameters with the lowest L1-norm \cite{paganini2020streamlining}, we observe that $\mathcal{F}$'s F1 significantly drops by 30.5\% on average. Except for the SST dataset, however, the Detection AUC still remains 93\% on average (Table \ref{tab:model_pruning}).

\renewcommand{\tabcolsep}{1.2pt}
\begin{table}[]
    \centering
    \small
    \begin{tabular}{ccccc}
  \toprule
  \textbf{Length} & \multicolumn{1}{c}{\textbf{50 words}} & \multicolumn{1}{c}{\textbf{100 words}} & \multicolumn{1}{c}{\textbf{250 words}} & \multicolumn{1}{c}{\textbf{500 words}} \\
 \cmidrule(lr){1-5}
  GF$\downarrow$ & 12 $\rightarrow$ 13 & 16$\rightarrow$17 & 23$\rightarrow$23 & 26$\rightarrow$26 \\
  \cmidrule(lr){1-5}
  Human$\uparrow$ & 7.5$\rightarrow$7.8 & 8.2$\rightarrow$7.5 & 7.4$\rightarrow$7.4 & 7.4$\rightarrow$7.0\\
  \bottomrule
    \end{tabular}
    \caption{Changes in average readability of varied-length news articles after UniTrigger attack using Gunning Fog (GF) score and human evaluation}
    \label{tab:readability}
\end{table}


\textbf{Parameters Analysis.} Regarding the trapdoor-ratio $\epsilon$, a large value (e.g., $\epsilon{\leftarrow}1.0$) can undesirably result in a detector network $\mathcal{G}$ that ``memorizes" the embedded trapdoors instead of learning its semantic meanings. A smaller value of $\epsilon{\leq}0.15$ generally works well across all experiments. Regarding the \textit{trapdoor weight} $\gamma$, while CNN and BERT are not sensitive to it, RNN prefers $\gamma{\leq}0.75$. Moreover, setting $\alpha$, $\beta$ properly to make them cover ${\geq}3000$ neighboring tokens is desirable.

\section{Related Work}

{\bf Adversarial Text Detection.}
%
Adversarial detection on NLP is rather limited. Most of the current detection-based adversarial text defensive methods focus on detecting typos, misspellings \cite{gao2018black,li2018textbugger,pruthi2019combating} or synonym substitutions \cite{wang2019natural}. Though there are several uncertainty-based adversarial detection methods \cite{smith2018understanding,sheikholeslami2020minimum,pang2018towards} that work well with computer vision, how effective they are on the NLP domain remains an open question.

\begin{table}[tb]
    \centering
    \small
    \begin{tabular}{cccccccccc}
    \toprule
    \multirow{2}{*}{\textbf{Pruning\%}} & \multicolumn{2}{c}{\textbf{MR}} & \multicolumn{2}{c}{\textbf{SJ}} & \multicolumn{2}{c}{\textbf{SST}} & \multicolumn{2}{c}{\textbf{AG}}  \\
    \cmidrule(lr){2-3}\cmidrule(lr){4-5}\cmidrule(lr){6-7}\cmidrule(lr){8-9}
    & \textbf{F1} & \textbf{AUC} & \textbf{F1} & \textbf{AUC} & \textbf{F1} & \textbf{AUC} & \textbf{F1} & \textbf{AUC} \\
    \midrule
    20\% & 64.9 & 99.3 & 80.0 & 99.2 & 37.3 & 68.2 & 17.1 & 98.5 \\
    50\% & 51.3 & 91.9 & 82.6 & 99.4  & 66.6 & 50.3 & 11.9 & 87.3 \\
    \bottomrule
    \end{tabular}
    \caption{\textbf{Model \underline{F1}} / \textbf{detect \underline{AUC}} of CNN under trapdoor removal using model-pruning}
    \label{tab:model_pruning}
\end{table}

\vspace{0.05in}
\noindent
{\bf Honeypot-based Adversarial Detection}
\cite{shan2019using} adopts the ``honeypot" concept to images. While this method, denoted as \textit{GCEA}, creates trapdoors via randomization, {\mymethod} generates trapdoors \textit{greedily}. Moreover, {\mymethod} only needs a single network $\mathcal{G}$ for adversarial detection. In contrast, GCEA records a separate neural signature (e.g., a neural activation pattern in the last layer) for each trapdoor. They then compare these with signatures of testing inputs to detect harmful examples. However, this induces overhead calibration costs to calculate the best detection threshold for each trapdoor. Furthermore, while \cite{shan2019using} and \cite{carlini2020partial} show that true trapdoors can be revealed and clustered by attackers after several queries on $\mathcal{F}$, this is not the case when we use {\mymethod} to defend against adaptive UniTrigger attacks (Sec. \ref{sec:adaptive}). Regardless of initial tokens (e.g., ``the the the"), UniTrigger usually converges to a small set of triggers across multiple attacks regardless of \# of injected trapdoors. Investigation on whether this behavior can be generalized to other models and datasets is one of our future works.

\section{Conclusion}
This paper proposes {\mymethod}, an algorithm that greedily injects multiple trapdoors, i.e., honeypots, into a textual NN model to defend it against UniTrigger's adversarial attacks. {\mymethod} achieves a TPR as high as 99\% and a FPR less than 2\% in most cases across four public datasets. We also show that {\mymethod} with more than one trapdoor is robust against even advanced attackers. While {\mymethod} only focuses on defending against UniTrigger, we plan to extend {\mymethod} to safeguard other NLP adversarial generators in future.

\bibliographystyle{acl_natbib}
\bibliography{acl2020}

\appendix
\clearpage
\newpage
\section{Appendix}
\label{sec:appendix}

\setcounter{table}{0}
\setcounter{figure}{0}
\renewcommand\thetable{\Alph{section}.\arabic{table}}
\renewcommand\thefigure{\Alph{section}.\arabic{figure}}

\subsection{Objective Function}\label{appendix:objective}
Eq. (\ref{eqn:objective}) details the full objective function of the \textit{Greedy Trapdoor Search} algorithm described in Sec. \ref{sec:multipletraps}.

\noindent
\vspace{-15pt}
\begin{center}
\fbox{\parbox[t]{0.95\linewidth}{
\textbf{\textsc{Objective Function 1}}: \textit{Given a NN $\mathcal{F}$, and hyper-parameter $\mathit{K}$, $\alpha$, $\beta$, our goal is to search for a set of $K$ trapdoors to defend each label $L \in \mathcal{C}$ by optimizing:}
\begin{equation}
\begin{aligned}
    &\minimize_{S^*_{L \in \mathcal{C}}}\;\sum_{L \in \mathcal{C}}\mathcal{L}^L_{\mathrm{fidelity}} \quad subject\;to\\
    &d(w_i, w_j) \leq \alpha\;\bigforall w_i, w_j \in S^*_L\\
    &d(w_i, w_j) \geq \beta\;\bigforall w_i \in S^*_L, w_j \in S^*_{Q \neq L}\\
    &L,Q \in \mathcal{C}, K \geq 1
\label{eqn:objective}
\end{aligned}
\end{equation}
\vspace{-10pt}
\label{objective}}}
\end{center}

\subsection{Further Details of  Experiments}\label{appendix:exp}
\begin{figure*}[t!]
  \centering
  \vspace{20pt}
  \includegraphics[width=0.95\textwidth]{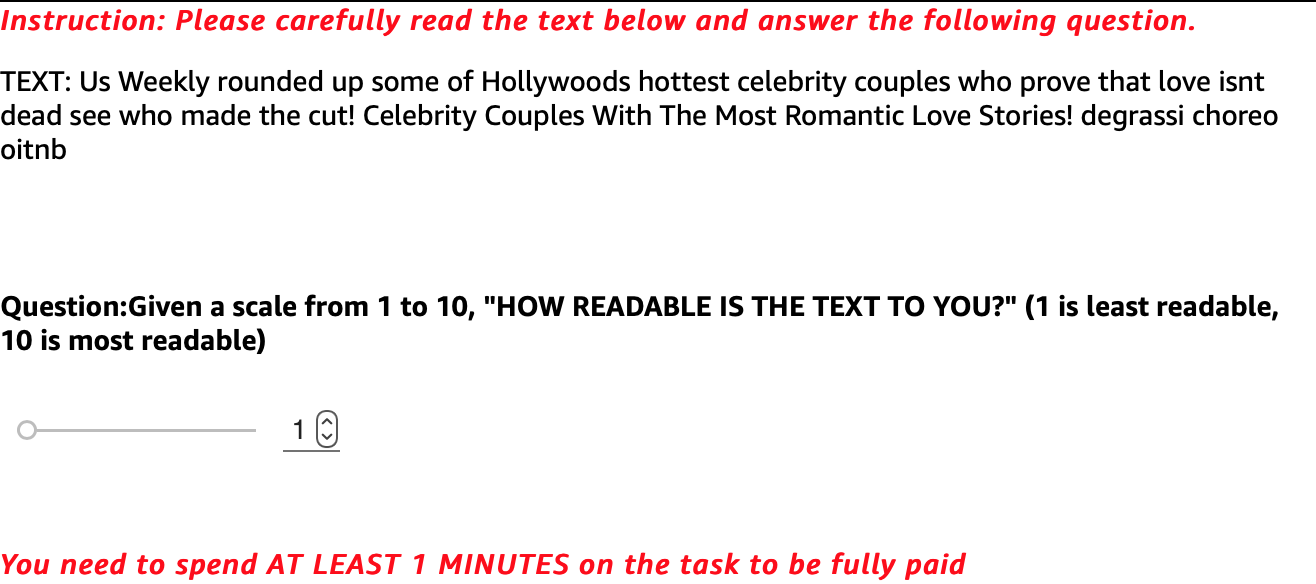}
  \caption{Example of user study interface for Sec. \ref{sec:discussion}}
  \label{fig:AMT}
\end{figure*}

\begin{figure}[t!]
  \centering
  \includegraphics[width=0.5\textwidth]{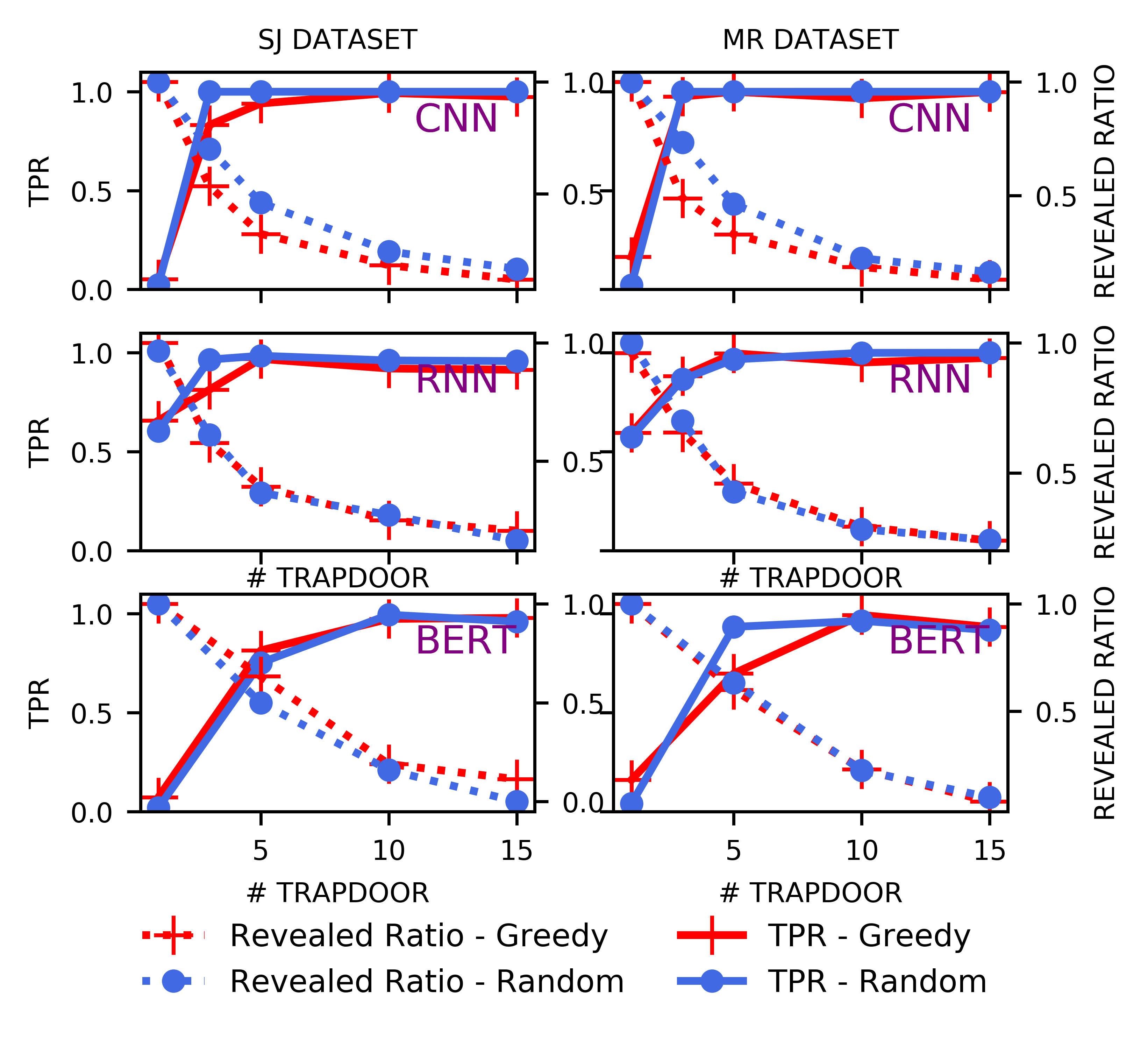}
  \caption{Performance under adaptive attacks}
  \label{fig:adaptive}
\end{figure}

\begin{figure}[t!]
  \centering
  \includegraphics[width=0.5\textwidth]{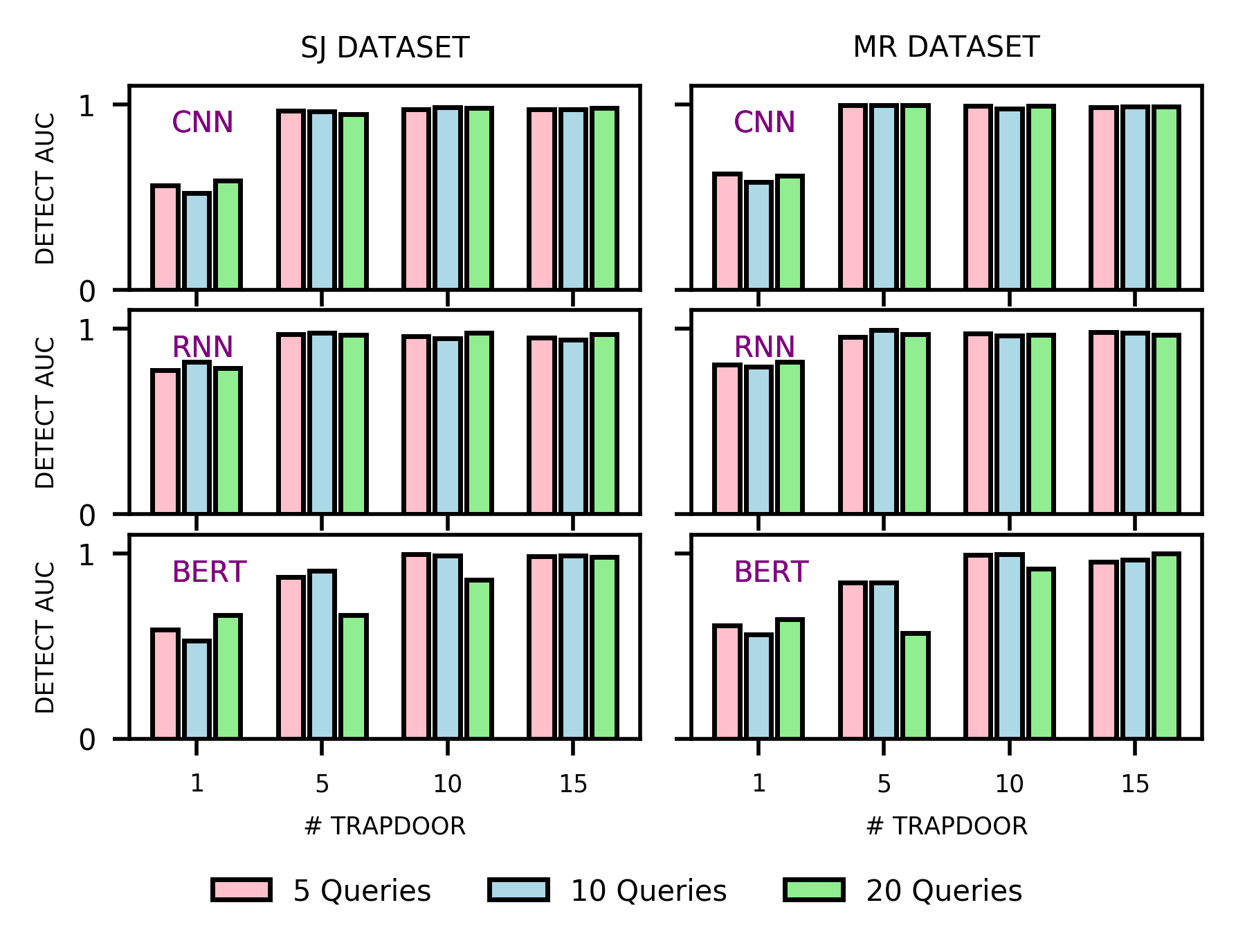}
  \caption{Detection AUC v.s. \# query attacks}
  \label{fig:query}
\end{figure}

\begin{itemize}[leftmargin=\dimexpr\parindent-0.2\labelwidth\relax,noitemsep,topsep=0pt]
    \item Table~\ref{tab:dataset} shows the detailed statistics of four datasets used in the experiments as mentioned in Sec. \ref{sec:setup}.
    \item Tables~\ref{tab:results_novice},~\ref{tab:results_whitebox_advanced},~\ref{tab:blackbox} show the performance results under the novice, advanced and black-box attack, respectively, as mentioned in Sec. \ref{sec:results}.

    \item Figure~\ref{fig:AMT} shows the user study design on Amazon Mechanical Turk as mentioned in Sec. \ref{sec:discussion}.
    \item Figures~\ref{fig:adaptive} and~\ref{fig:query} show the performance under the adaptive attack as mentioned in Sec. \ref{sec:results}.  
    
    \end{itemize}

\renewcommand{\tabcolsep}{6.2pt}
\begin{table*}[tb]
 \centering
 \small
 \begin{tabular}{lcccccc}
\toprule
\textbf{Dataset} & \textbf{Acronym} & \textbf{\# Class} & \textbf{Vocabulary Size} & \textbf{\# Words} & \textbf{\# Data} \\
\midrule
Subjectivity & SJ & 2 & 20K & 24 & 10K \\
Movie Reviews &MR & 2 & 19K & 21 & 11K \\
Sentiment Treebank &SST & 2 & 16K & 19 & 101K \\
AG News &AG & 4 & 71K & 38 & 120K \\
\bottomrule
 \end{tabular}
\caption{Dataset statistics}
 \label{tab:dataset}
\end{table*}

\renewcommand{\tabcolsep}{4pt}
\begin{table*}[t!b]
\centering
\small
\vspace{20pt}
\begin{tabular}{clcccccccccccc}
\toprule
\multirow{3}{*}{\textbf{}} & \multicolumn{1}{c}{\multirow{3}{*}{\textbf{Method}}} & \multicolumn{4}{c}{\textbf{RNN}} & \multicolumn{4}{c}{\textbf{CNN}} & \multicolumn{4}{c}{\textbf{BERT}}  \\
\cmidrule(lr){3-14}
 &  & \multicolumn{1}{c}{\textbf{Clean}} & \multicolumn{3}{c}{\textbf{Detection}} & \multicolumn{1}{c}{\textbf{Clean}} & \multicolumn{3}{c}{\textbf{Detection}} & \multicolumn{1}{c}{\textbf{Clean}} & \multicolumn{3}{c}{\textbf{Detection}}\\
\cmidrule(lr){4-6}\cmidrule(lr){8-10}\cmidrule(lr){11-14}
{} &  & \textbf{F1} & \textbf{AUC} & \multicolumn{1}{c}{\textbf{FPR}} & \multicolumn{1}{c}{\textbf{TPR}}& \textbf{F1} & \textbf{AUC} & \multicolumn{1}{c}{\textbf{FPR}} & \multicolumn{1}{c}{\textbf{TPR}} & \textbf{F1} & \textbf{AUC} & \multicolumn{1}{c}{\textbf{FPR}} & \multicolumn{1}{c}{\textbf{TPR}} \\
\midrule
 & OOD & \underline{76.5} & 47.3 & 49.0 & 51.0 & \textbf{78.9} & 82.3 & 23.5 & 78.4 & \underline{84.7} & 38.4 &  61.3 &  50.7\\
 & ScRNN & - & 55.1 & 43.1 & 53.7 & - & 54.7 & 43.1 & 53.1 & - & 52.0 &  52.3 &  55.1\\
M & USE & - & 64.8 & 46.1 & 77.7 & - & 64.8 & 45.3 & 74.6 & - & 49.5 &  57.3 &  60.7 \\
R & SelfATK & - & 96.5 &  \underline{0.8} &  93.9 & - & 97.0 &  \underline{0.1} &  94.1 & - & \underline{93.4} &  \underline{4.0} &  \underline{87.5} \\
& LID & - & 53.2 &  44.1 &  50.6 & - & 66.2 &  42.5 &  74.9 & - & 55.4 &  51.5 &  61.9 \\
\cmidrule(lr){2-14}
 & {\mymethod}(1) & 75.9 & \textbf{99.9} & \textbf{0.2} & \textbf{100.0} & 74.6 & \underline{98.4} & \textbf{0.5} & \underline{97.3} & \textbf{85.0} & 91.7 &  \textbf{3.9} &  84.0 \\
 & {\mymethod}(5) & \textbf{78.0} & \underline{99.1} & 1.0 & \underline{99.5} & \underline{77.3} & \textbf{99.4} & 1.1 & \textbf{100.0} & 84.2 & \textbf{100.0} &  \underline{4.0} &  \textbf{100.0}\\
\midrule
 & OOD & 88.5 & 34.3 & 64.9 & 47.1 & \textbf{90.1} & 82.6 & 23.6 & 79.9 & 95.8 & 20.9 &  76.3 &  42.1\\
 & ScRNN & - & 53.6 & 47.8 & 55.6 & - & 59.8 & 43.9 & 59.7 & - & 53.4 &  53.6 &  58.6\\
S & USE & - & 65.2 & 45.2 & 77.0 & - & 74.6 & 37.5 & 83.8 & - & 62.5 &  50.8 &  75.7\\
J & SelfATK & - & \underline{98.5} &  1.9 &  \underline{98.9} & - & \underline{98.5} &  \underline{0.1} &  \underline{97.1} & - & \underline{98.8} &  \underline{6.2} &  \underline{97.9}\\
& LID & - & 48.9 &  53.0 &  50.8 & - & 71.7 &  29.2 &  72.7 & - & 61.9 &  56.0 &  78.4\\
\cmidrule(lr){2-14}
 & {\mymethod}(1) & \underline{89.5} & \textbf{99.5} & \textbf{0.3} & \textbf{99.2} & 88.1 & 97.6 & \textbf{0.8} & 95.9 & \textbf{96.1} & \textbf{100.0} &  \textbf{6.1} &  \textbf{100.0}\\
 & {\mymethod}(5) & \textbf{89.8} & 97.4 & \underline{1.2} & 96.0 & \underline{89.6} & \textbf{99.2} & 1.5 & \textbf{100.0} & \underline{96.0} & \textbf{100.0} &  \underline{6.2} &  \textbf{100.0}\\
\midrule
 & OOD & \textbf{84.4} & 50.8 & 47.3 & 51.8 & \textbf{81.1} & 86.1 & 19.4 & 81.6 & 93.5 & 33.3 &  63.6 &  43.4 \\
 & ScRNN & - & 54.4 & 19.1 & 27.8 & - & 55.1 & 19.1 & 29.3 & - & 50.2 &  50.6 &  51.2 \\
S & USE & - & 58.1 & 51.3 & 68.7 & - & 51.0 & 58.5 & 67.8 & - & 55.7 &  51.2 &  62.6 \\
S & SelfATK & - & 67.1 &  \underline{2.9} &  37.1 & - & 83.8 &  \textbf{0.2} &  67.8 & - & 82.6 &  \textbf{1.6} &  65.7 \\
T & LID & - & 50.0 &  41.3 &  41.3 & - & 71.1 &  20.9 &  63.2 & - & 48.6 &  \underline{43.8} &  40.9\\
\cmidrule(lr){2-14}
 & {\mymethod}(1) & \underline{83.5} & \underline{96.6} & 6.8 & \underline{99.9} & 77.4 & \underline{98.1} & \underline{0.4} & \underline{96.7} & \textbf{94.2} & \underline{91.6} &  \textbf{1.6} &  \underline{83.6}\\
 & {\mymethod}(5) & 82.6 & \textbf{99.6} & \textbf{0.8} & \textbf{100.0} & \underline{79.3} & \textbf{98.5} & 2.4 & \textbf{99.3} & \underline{93.9} & \textbf{100.0} &  \textbf{1.6} &  \textbf{100.0}\\
\midrule
 & OOD & \textbf{91.0} & 44.4 & 51.5 & 47.7 & \textbf{89.6} & 67.3 & 34.7 & 61.9 & 93.2 & 27.5 &  69.8 &  41.9\\
 & ScRNN & - & 53.1 & 48.4  & 52.9 & - & 53.6 & 47.7 & 52.8 & - & 51.7 &  \underline{50.6} &  53.2\\
A & USE & - & 81.6 & 29.6 & 86.9 & - & 67.2 & 44.0 & 78.1 & - & 57.6 &  52.8 &  70.0\\
G & SelfATK & - & 92.6 &  \textbf{4.3} &  89.5 & - & 93.2 &  \underline{3.9} &  90.4 & - & \textbf{99.8} &  \textbf{0.1} &  \textbf{99.6} \\
& +LID & - & 55.5 &  45.3 &  56.3 & - & 79.8 &  23.1 &  82.6 & - & 48.5 &  54.7 &  51.6\\
\cmidrule(lr){2-14}
 & {\mymethod}(1) & 89.7 & \textbf{97.2} & \underline{5.4} & \textbf{99.8} & 88.2 & \textbf{98.9} & \textbf{2.0} & \underline{99.7} & \textbf{93.9} & 89.3 &  \textbf{0.1} &  78.7 \\
& {\mymethod}(5) & \underline{89.9} & \underline{96.5} & 6.8 & \textbf{99.8} & \underline{88.8} & \underline{94.5} & 11.0 & \textbf{100.0} & \underline{93.3} & \underline{97.6} &  \textbf{0.1} &  \underline{95.4}\\
\bottomrule
\end{tabular}
\caption{Average detection performance across all target labels under novice attack}
\label{tab:results_novice}
\end{table*}

\renewcommand{\tabcolsep}{4pt}
\begin{table*}[t!b]
\centering
\small
\begin{tabular}{clcccccccccccc}
\toprule
\multirow{3}{*}{\textbf{}} & \multicolumn{1}{c}{\multirow{3}{*}{\textbf{Method}}} & \multicolumn{4}{c}{\textbf{RNN}} & \multicolumn{4}{c}{\textbf{CNN}} & \multicolumn{4}{c}{\textbf{BERT}}  \\
\cmidrule(lr){3-14}
 &  & \multicolumn{1}{c}{\textbf{Clean}} & \multicolumn{3}{c}{\textbf{Detection}} & \multicolumn{1}{c}{\textbf{Clean}} & \multicolumn{3}{c}{\textbf{Detection}} & \multicolumn{1}{c}{\textbf{Clean}} & \multicolumn{3}{c}{\textbf{Detection}}\\
\cmidrule(lr){4-6}\cmidrule(lr){8-10}\cmidrule(lr){11-14}
{} &  & \textbf{F1} & \textbf{AUC} & \multicolumn{1}{c}{\textbf{FPR}} & \multicolumn{1}{c}{\textbf{TPR}}& \textbf{F1} & \textbf{AUC} & \multicolumn{1}{c}{\textbf{FPR}} & \multicolumn{1}{c}{\textbf{TPR}} & \textbf{F1} & \textbf{AUC} & \multicolumn{1}{c}{\textbf{FPR}} & \multicolumn{1}{c}{\textbf{TPR}} \\
\toprule
 & OOD & 75.2 & 52.5 & 45.9 & 55.7 & \textbf{77.7} & \underline{74.8} &  30.0 &  72.4 & \textbf{84.7} & 35.6 &  63.9 &  48.2\\
 & ScRNN & - & 51.9 &  43.0 &  47.0 & - & 57.3 &  41.6 &  56.4 & - & 51.8 &  52.3 &  54.9 \\
M & USE & - &  62.9 &  48.1 &  75.9 &  - & 66.2 &  44.5 &  \underline{77.7} & - & 53.1 &  55.1 &  64.1  \\
R & SelfATK & - & \textbf{92.3} &  \underline{0.6} &  \underline{85.1}  & - & 69.8 &  \textbf{0.4} &  40.0 & - & \textbf{97.5} &  4.1 &  \textbf{95.2}\\
& LID & - & 51.3 &  45.8 &  48.4  & - & 66.2 &  37.4 &  69.7 & - & 54.2 &  51.5 &  59.6\\
\cmidrule(lr){2-14}
 & {\mymethod}(1) & \underline{77.8} & 74.8 &  \textbf{0.8} &  50.4 & 76.9 & 73.6 &  \textbf{0.4} &  47. & \textbf{84.7} & 74.3 &  \textbf{3.9} &  50.7\\
 & {\mymethod}(5) & \textbf{78.1} & \textbf{92.3} &  2.9 &  \textbf{87.6} & \underline{77.4} & \textbf{91.2} &  \underline{3.2} &  \textbf{85.5} & \underline{84.3} & \underline{92.3} &  \underline{4.0} &  \underline{85.3} \\
\toprule
 & OOD & \textbf{89.4} & 34.5 & 62.5 & 43.1  & \textbf{89.6} & 59.9 &  44.2 &  64.7 & \underline{96.1} & 21.9 &  74.6 &  43.6\\
 & ScRNN & - & 57.6 &  51.1 &  65.7 & - & 55.0 &  53.6 &  62.9 & - & 53.1 &  53.6 &  58.1\\
S & USE & - & 70.7 &  41.4 &  \underline{81.6} & - & 72.7 &  38.8 &  \underline{83.1} & - & 65.7 &  48.5 &  74.4 \\
J & SelfATK & - & \underline{80.7} &  8.0 &  69.3 & - & \underline{72.8} &  \textbf{0.5} &  46.0 & - & \underline{96.8} &  \underline{6.2} &  \underline{94.0} \\
& LID & - & 50.7 &  54.3 &  55.7 & - & 67.5 &  32.0 &  67.1 & - & 62.2 &  56.1 &  79.0\\
\cmidrule(lr){2-14}
 & {\mymethod}(1) & \textbf{89.4} &  71.7 &  \textbf{0.6} &  43.9 & \underline{88.5} & 70.8 &  4.9 &  46.6 & \textbf{96.2} & 68.6 &  \textbf{6.1} &  41.0 \\
 & {\mymethod}(5) & \underline{88.9} & \textbf{92.7} &  \underline{2.4} &  \textbf{87.9} & 87.6 & \textbf{93.9} &  \underline{4.3} &  \textbf{92.0} & \underline{96.1} & \textbf{100.0} &  \underline{6.2} &  \textbf{100.0} \\
\toprule
 & OOD & 79.0 & 50.6 & 48.8 & 52.5  & \underline{77.7} & \underline{77.7} &  26.3 &  74.2 & 93.6 & 31.3 &  67.1 &  45.7\\
 & ScRNN & - & 53.8 &  19.2 &  26.8 & - & 56.1 &  19.1 &  31.2 & - & 53.2 &  50.3 &  54.9 \\
S & USE & - & 60.8 &  50.1 &  \underline{72.2} & - & 55.2 &  55.4 &  \underline{70.4} & - & 51.0 &  57.7 &  63.7\\
S & SelfATK & - & 66.1 &  3.7 &  35.9 & - & 61.8 &  \textbf{0.2} &  23.8 & - & \underline{91.1} &  \underline{1.7} &  \underline{82.5}\\
T & LID & - & 49.9 &  62.2 &  61.9 & - & 64.0 &  18.8 &  46.9 & - & 46.2 &  42.6 &  35.1 \\
\cmidrule(lr){2-14}
 & {\mymethod}(1) & \underline{82.9} & \underline{69.7} &  \textbf{0.2} &  39.6 & 77.3 & 59.3 &  \underline{0.9} &  19.6 & \textbf{94.2} & 50.0 &  \textbf{1.6} &  1.6\\
 & {\mymethod}(5) & \textbf{83.3} & \textbf{93.1} &  \underline{3.2} &  \textbf{89.4}  & \textbf{78.7} & \textbf{83.0} &  5.4 &  \textbf{71.5} & \underline{94.1} & \textbf{94.6} &  \textbf{1.6} &  \textbf{89.4} \\
\toprule
 & OOD & \textbf{90.9} & 40.5 & 56.3 & 46.9  &  \textbf{89.4} & 63.1 &  38.2 &  59.0 & 93.1 & 26.9 &  69.2 &  40.7 \\
 & ScRNN & - & 56.0 &  46.1 &  54.7 & - & 53.7 &  48.8 &  54.1 & - & 54.4 &  \underline{46.4} &  52.6 \\
A & USE & - &  \underline{88.6} &  22.7 &  \underline{90.5} & - & 69.4 &  42.0 &  78.7 & - & 60.0 &  50.3 &  70.8\\
G & SelfATK & - & 88.4 &  \textbf{6.2} &  83.1 & - & 80.7 &  \textbf{8.0} &  69.4 & - & \underline{92.0} &  \textbf{0.1} &  \underline{84.0} \\
& LID & - & 54.3 &  45.9 &  54.6  & - & 79.1 &  22.1 &  80.3 & - & 48.3 &  52.9 &  49.4\\
\cmidrule(lr){2-14}
 & {\mymethod}(1) & 87.4 & 54.0 &  80.4 &  88.4 & 86.6 & \underline{83.3} &  19.0 &  85.5 & \textbf{93.9} & 70.3 &  \textbf{0.1} &  40.7 \\
& {\mymethod}(5) & \underline{89.7} & \textbf{95.2} &  \underline{9.3} &  \textbf{99.8} & \underline{88.6} & \textbf{92.6} &  \underline{14.7} &  \textbf{99.9} & \underline{93.3} & \textbf{97.0} &  \textbf{0.1} &  \textbf{94.0}\\
\bottomrule
\end{tabular}
\caption{Average detection performance across all target labels under advanced attack}
\label{tab:results_whitebox_advanced}
\end{table*}

\subsection{Reproducibility}
\subsubsection{Source Code} We will release the source code of {\mymethod} upon the acceptance of this paper. 

\subsubsection{Computing Infrastructure} We run all experiments on the machines with Ubuntu OS (v18.04), 20-Core Intel(R) Xeon(R) Silver 4114 CPU @ 2.20GHz, 93GB of RAM and a Titan Xp GPU. All implementations are written in Python (v3.7) with Pytorch (v1.5.1), Numpy (v1.19.1), Scikit-learn (v0.21.3). We also use the \textit{Transformers} (v3.0.2)\footnote{\url{https://huggingface.co/transformers/}}
library for training \textit{transformers-based} BERT.


\renewcommand{\tabcolsep}{5pt}
\begin{table}[t!b]
\centering
\vspace{5pt}
\small
\begin{tabular}{ccccc}
\toprule
\multirow{3}{*}{}  & \multicolumn{2}{c}{\textbf{Adaptive}} & \multicolumn{2}{c}{\textbf{Random}} \\
\cmidrule(lr){2-3}\cmidrule(lr){4-5}
& \textbf{Detect} & \textbf{Attack} & \textbf{Detect} & \textbf{Attack} \\
 & \textbf{AUC}$\uparrow$ & \textbf{ACC}$\downarrow$ & \textbf{AUC}$\uparrow$ & \textbf{ACC}$\downarrow$ \\
\midrule
MR & 74.24 & 4.6 & 85.3 & 3.77\\
SJ & 87.19 & 0.34 & 76.78 & 2.86\\
SST & \textcolor{red}{58.81} & 19.77 & \textcolor{red}{49.75} & 18.96 \\
AG & 67.88 & 55.87 & \textcolor{red}{53.25} & 75.25\\
\bottomrule
\multicolumn{5}{l}{\textcolor{red}{Red}: not transferable}\\
\end{tabular}
\caption{Detection AUC and model's accuracy (attack ACC) under black-box attack on CNN}
\label{tab:blackbox}
\end{table}
\subsubsection{Average Runtime} According to Sec. \ref{sec:framework}, the computational complexity of greedy trapdoor search scales linearly with the number of labels $|\mathcal{C}|$ and vocabulary size $|\mathcal{V}|$. Moreover, the time to train a detection network depends on the size of a specific dataset, the trapdoor ratio $\epsilon$, and the number of trapdoors $K$. 

For example, {\mymethod} takes roughly 14 and 96 seconds to search for 5 trapdoors to defend each label for a dataset with 2 labels and a vocabulary size of 19K (e.g., Movie Reviews) and a dataset with 4 labels and a vocabulary size of 91K (e.g., AG News), respectively. With $K{\leftarrow}5$ and $\epsilon{\leftarrow}0.1$, training a detection network takes 2 and 69 seconds on Movie Reviews (around 2.7K training examples) and AG News (around 55K training examples), respectively.

\subsubsection{Model's Architecture and \# of Parameters} 
The \textit{CNN} text classification model  with 6M parameters~\cite{kim2014convolutional}   has three 2D convolutional layers (i.e., 150 kernels each with a size of 2, 3, 4) followed by a \textit{max-pooling} layer, a dropout layer with 0.5 probability, and a fully-connected-network (FCN) with softmax activation for prediction. We use the pre-trained \textit{GloVe}~\cite{pennington2014glove}
embedding layer of size 300 to transform each discrete text tokens into continuous input features before feeding them into the model. The \textit{RNN}  text model with 6.1M parameters replaces the convolution layers of CNN with a GRU network of 1 hidden layer. The \textit{BERT} model with 109M parameters is imported from the transformers library. We use the \textit{bert-base-uncased} version of BERT. 

\subsubsection{Hyper-Parameters} Sec. \ref{sec:discussion} already discussed the effects of all hyper-parameters on {\mymethod}'s performance  as well as the most desirable values for each of them. To tune these hyper-parameters, we use the grid search as follows: $\epsilon \in \{1.0, 0.5, 0.25, 0.1\}$, $\gamma \in \{1.0, 0.75, 0.5\}$. Since $\alpha$ and $\beta$ are sensitive to the domain of the pre-trained word-embedding (we use GloVe embeddings~\cite{pennington2014glove}), without loss of generality, we instead use \# of neighboring tokens to accept or filter to search for the corresponding $\alpha, \beta$ in Eq. (\ref{eqn:classawareness}): $\{500, 1000, 3000, 5000\}$. 

We set the number of randomly sampled candidate trapdoors to around 10\% of the vocabulary size ($T{\leftarrow}300$). We train all models using a learning rate of 0.005 and batch size of 32. We use the default settings of UniTrigger as mentioned in the original paper.

\subsubsection{Datasets} We use \textit{Datasets} (v1.2.1)\footnote{\url{https://huggingface.co/docs/datasets/}}
library to load all the standard benchmark datasets used in the paper, all of which are publicly available.
\end{document}